\documentclass[sigconf]{acmart}

\AtBeginDocument{%
  \providecommand\BibTeX{{%
    \normalfont B\k ern-0.5em{\scshape i\kern-0.25em b}\kern-0.8em\TeX}}}

\usepackage{booktabs} 
\usepackage{epigraph}
\usepackage{algorithm}
\usepackage{algpseudocode}
\usepackage{graphicx}
\usepackage{multirow}

\usepackage{amssymb}
\usepackage{amsmath}
\usepackage{amsfonts}
\usepackage{color}
\usepackage{url}
\usepackage{subfigure}
\usepackage{todonotes}
\usepackage[shortlabels]{enumitem}
\usepackage{blindtext}
\usepackage{hyperref}
\def\0{{\bf 0}}
\def\1{{\bf 1}}

\def\AM{{\mathcal A}}

\def\HM{{\mathcal H}}
\def\IM{{\mathcal I}}

\def\RM{{\mathcal R}}

\def\VM{{\mathcal V}}

\newcommand{\nop}[1]{}

\newtheorem{definition}{Definition}

\definecolor{Gray}{gray}{0.9}

\begin{document}

\title{Kuaipedia: a Large-scale Multi-modal Short-video Encyclopedia}

\author{Haojie Pan$^1$, Zepeng Zhai$^1$, Yuzhou Zhang$^{2}$, Ruiji Fu$^{1\dagger}$, Ming Liu $^2$, \\ Yangqiu Song$^3$, Zhongyuan Wang$^1$, Bing Qin$^2$}

\affiliation{$^1$ Kuaishou Inc. $^2$ Harbin Institute of Technology $^3$ HKUST
\institution{\{panhaojie,zhaizepeng03,furuiji,wangzhongyuan\}@kuaishou.com}
\country{\{yuzhouzhang, mliu, qinb\}@ir.hit.edu.cn, \{yqsong\}@cse.ust.hk}
\city{~ \\ ~ \\ ~ \\ ~}
}
\begin{abstract}
The rapid growth of online encyclopedias, such as Wikipedia, has revolutionized the way people access and share information. 
However, the traditional text, images, and tables can hardly express some aspects of a wiki item. For example, when we talk about the dog breed ``Shiba Inu'', one may care more about ``How to feed it'' or ``How to train it not to protect its food''. 
Short-video platforms, such as TikTok, Kuaishou, and YouTube Shorts, have become a hallmark in the online world and are popular sources for sharing knowledge and insights on a wide range of topics.
Those knowledge-sharing videos provide a concise and visually appealing way to convey information about a particular item, such as hair characteristics or feeding instructions of a ``Shiba Inu'', which can be efficiently analyzed and organized in a manner similar to an online encyclopedia.
In this paper, we propose Kuaipedia, a large-scale multi-modal encyclopedia consisting of items, aspects, and short videos lined with them, sourced from billions of videos of Kuaishou (Kwai), a well-known short-video platform in China. 
We first collected items from multiple sources and mined user-centered aspects from millions of users' queries to build item-aspect trees. Then we propose a new task called ``multi-modal item-aspect linking'' as an expansion of ``entity linking'' to ground short videos into item-aspect pairs and build the whole short-video encyclopedia. 
Intrinsic evaluations show that our encyclopedia is of large scale and highly accurate. 
We have conducted extensive extrinsic evaluations to demonstrate the effectiveness of Kuaipedia in enhancing fundamental applications such as entity typing and linking.  Moreover, our findings show that the multi-modal information in Kuaipedia can enhance the professionality and factual accuracy of language models such as ChatGPT and Dall·E.
\footnote{~~The data and experimental results will be released on the homepage of this paper.}

\end{abstract}



\maketitle
\section{Introduction}







Encyclopedia, dating back to ancient Greek and Roman civilizations, was further developed during the French Enlightenment in the 17th and 18th centuries.  It serves as a comprehensive reference compendium, providing summaries of knowledge across various fields and aspects. Under the thriving development of the Internet, there comes online encyclopedia such as Wikipedia~\citep{wikipedia}, BaiduBaike~\citep{baidubaike} for general knowledge and Investopedia~\citep{investopedia} for domain-specific knowledge. These digital encyclopedias offer a rich tapestry of information, combining text, images, and structured tables to present a complete picture of a given topic in a single article.

\begin{figure}[t]
\centering
\includegraphics[height=0.4\textwidth]{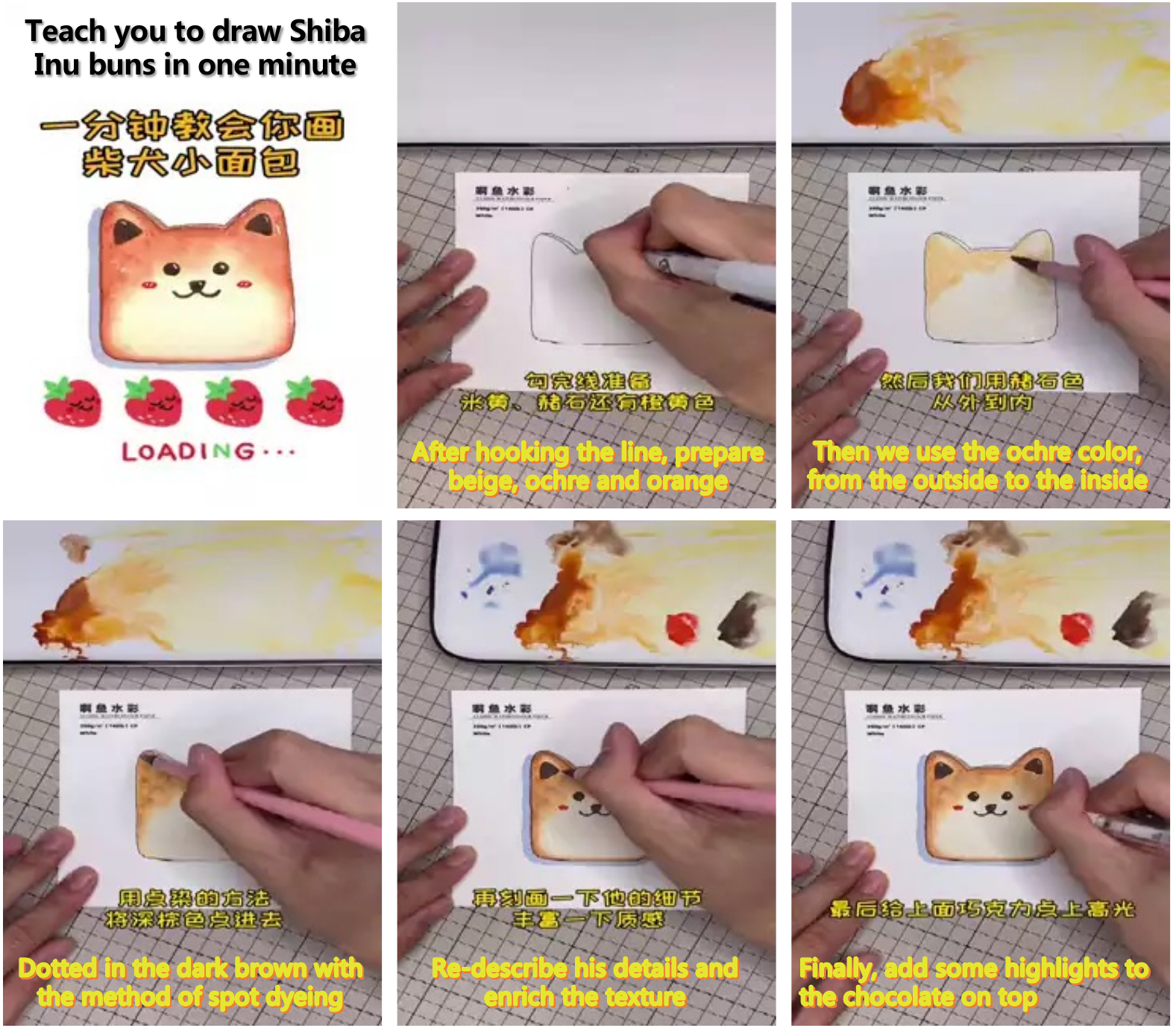}
\caption{A case of key frames of a video that explains how to draw a cartoon image for Shiba Inu buns. It is difficult for us to learn how-to-draw only by using pictures and text. }\label{fig:howtocase}
\end{figure}

On the other hand, the resurgence of knowledge engineering in recent years provides many knowledge graphs (KG) for encyclopedia knowledge (e.g. Freebase~\citep{DBLP:conf/sigmod/BollackerEPST08}, DBpedia~\citep{journals/semweb/LehmannIJJKMHMK15}, YAGO~\citep{suchanek2007semantic}, WikiData~\citep{VrandecicKroetzsch14cacm}, CN-Dbpedia ~\citep{DBLP:conf/ieaaie/XuXLXLCX17}). 
As KGs with pure symbols denoted in the form of text weaken the machines' capability of understanding the world~\citep{DBLP:journals/corr/abs-2202-05786}, researchers proposed many Multi-Modal KG(MMKG)s such as NEIL~\citep{DBLP:conf/iccv/ChenSG13}, IMGpedia~\citep{DBLP:conf/semweb/FerradaBH17} and Richpedia~\citep{DBLP:journals/bdr/WangWQZ20}, etc.
Those encyclopedias, KGs and MMKGs, which mostly depend on their text, images, and tables, suffer from describing knowledge that needs to be shown alive (e.g. ``how-to'' knowledge). Figure \ref{fig:howtocase} shows the difficulty for people to learn ``how-to'' knowledge only by the usage of text and pictures. However, we can find it's easier to learn by videos \footnote{~~The original video in Figure \ref{fig:howtocase} can be found in \url{https://www.gifshow.com/fw/photo/3xhcmzgr9fq492m}.}. That spatial and temporal information, or \textit{script knowledge}~\citep{DBLP:conf/ijcai/SchenkA75}, inside a video is important for machines to understand the world and is the key features of the capability for commonsense reasoning~\citep{DBLP:conf/nips/ZellersLHYPCFC21}.

In recent years, short videos, which do not exceed five or ten minutes in duration, have sprung up on the Internet and have become a trending form to gain new information and knowledge while sharing different skills and crafts ~\citep{Zhang2020}. Platforms such as TikTok, Kuaishou (Kwai), Instagram, or YouTube Shorts show the relative convenience of content generation and rapid content transmission. Existing works such as the website \textit{\url{check123.com}} or 
\textit{\url{baike.baidu.com}} show the considerable potential to use short videos to explain any knowledge in the world. Most of the short videos on these websites are used to explain an introduction or ``know-what'' knowledge of items (e.g. a brief introduction of Shiba Inu), which underestimates the representation power of the short videos. Like the video shown in Figure \ref{fig:howtocase}, there are also plentiful short videos to explain ``know-how'' or ``know-why'' knowledge.
Furthermore, the ``introduction'' is just the tip of the iceberg when it comes to knowledge videos about Shiba Inu. In addition to the ``introduction'', we may delve into more interesting aspects such as the breed's ``temperament'', ``price'', ``handshake'', and ``food-protection''. These topics cannot be effectively explained in a single short video, making it imperative to fully utilize these videos by exploring each topic in depth.

\begin{figure}[t]
\centering
\includegraphics[height=0.35\textwidth]{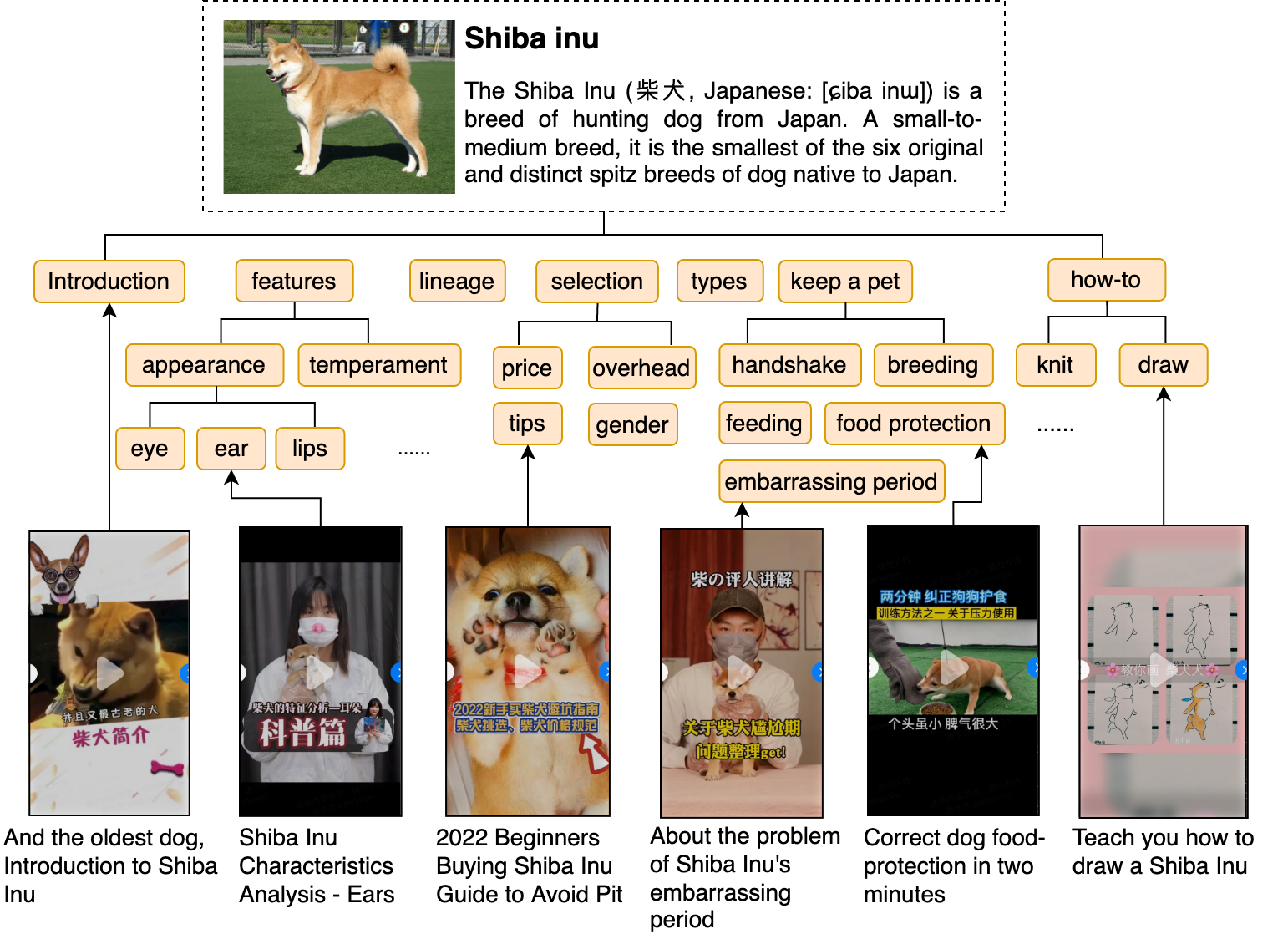}
\caption{An example of Kuaipedia. Aspects and linked viedos of Item ``Shiba Inu''. As for one aspect, there are multiple videos linked to it.}\label{fig:overview}
\vspace{-1.5pt}
\end{figure}

Here we propose Kuaipedia, the first structured multi-modal short-video encyclopedia in the world, which explains aspects of items by using short videos. Items, aspects, and videos are the three main elements in Kuaipedia. \textit{Items} is a set of entities and concepts, such as ``Shiba Inu'', ``Dog''; \textit{Aspects} is a set of keywords or keyphrases about the items, such as ``temperament'' to ``Shiba Inu''; Videos is a set of short videos that provide knowledge about specific aspects of the items. One comprehensive example of a Kuaipedia page is shown in Figure \ref{fig:overview}. 

Kuaipedia was extracted from billions of short videos on Kuaishou, one of the most famous short-video platforms in China. We first trained a knowledge video detection model to filter about 200 million knowledge videos. Then we collect more than 20 million items from multi-sources and design an item-aspect mining pipeline to extract more than 70 million item-aspect pairs. After that, we propose a new task called ``multi-modal item-aspect linking'', which extends the ``entity linking'' task. This task identifies item mentions in the short videos, links them to Kuaipedia items, and then utilizes a BERT-based ranking module to select the most relevant item-aspect pair for the video. Our intrinsic evaluations reveal that (1) Kuaipedia has competitive scalability in terms of the number of items, aspects, and videos, and (2) the mined aspects and linked video-item-aspect pairs have high quality and accuracy.
Extensive extrinsic experiments demonstrate the benefits of incorporating multi-modal knowledge from Kuaipedia into downstream applications such as entity typing and linking. Our research also demonstrates that the integration of multi-modal information within Kuaipedia can significantly improve the professional tone and factual accuracy of advanced language models, such as ChatGPT~\citep{Brown2020LanguageModels} and Dall·E~\cite{DBLP:journals/corr/abs-2204-06125}. 

The contributions of the paper conclude as follows:

1. \textbf{Definition of Kuaipedia}. We define a brand-new multi-modal encyclopedia where the primitive units are items, aspects, and short videos. It is the first structured short-video encyclopedia organized by items and aspects.

2. \textbf{Scalable Extraction of Kuaipedia}. We perform knowledge video detection, item-aspect mining, and multi-modal item-aspect linking over large-scale short videos. The latter is an extension of traditional ``entity linking'' task.

3. \textbf{Evaluations of Kuaipedia}. We thoroughly evaluate Kuaipedia's quality and effectiveness through experiments and human annotations. The results of our experiments in various applications, including entity typing, entity linking, and language model prompting, demonstrate the promising potential of Kuaipedia as a multi-modal encyclopedia.
\section{Overview of Kuaipedia}

Kuaipedia consists of \textit{items}, \textit{aspects}, \textit{videos} and their \textit{relations}, which differs from traditional knowledge graphs. Thus we devise the formal definition of Kuaipedia as below.

\begin{definition}
{\bf Kuaipedia} is a multi-modal hybrid graph $\HM$ of video items $\IM$'s, aspects $\AM$'s, videos $\VM$'s, and their relations $\RM$'s. Each {\bf item} $I$ is either an entity or a concept that can be found on a wiki page. Each {\bf aspect} $A$ is either a keyword or a keyphrase that has meanings of one aspect of an item. Each {\bf video} $V$ consists of its raw frame features and other machine-generated features. We also define three types of relations. $R_1$ over $\{A_i, I_j \}$ refers the aspect $A_i$ is belonging to the item $I_j$, $R_2$ over $ \{A_i, A_j\}$ refers the aspect $A_i$ is a hyponyms of aspect $A_j$ , and $R_3$ over $\{V_i, I_j, A_k \}$ means the main content of video $V_i$ is about the aspect $A_k$ of $I_j$. And the {\bf relation set} $\RM = \{ R_1, R_2, R_3 \}$. Overall, we have Kuaipedia $\HM=\{\VM, \IM, \AM, \RM\}$.
\end{definition}

\begin{definition}
{\bf Item-aspect trees (IAT)} is Kuaipedia excluding videos, and we denote it as $\HM' = \{\IM, \AM, \RM'\}$, where $\RM'=\{ R_1, R_2\}$

\end{definition}

A detailed explanations of  \textit{videos}, \textit{items}, \textit{aspects} are defined as follows:

\noindent $\bullet$ \textbf{\textit{Items}} is a set of entities and concepts, such as ``Shiba Inu'', ``Moon'', ``Galileo Galilei'', which can be edited at one Wikipedia page. An item may have a title, a subtitle, a summary, attributes, and other detailed information of the item.

\noindent $\bullet$  \textbf{\textit{Aspects}} is a set of keywords or keyphrases attached to items. Those keywords are used to describe specific aspects of the item. For example, ``selection'', ``food-protecting'', ``color'' of item ``Shiba Inu'', or ``formation'', ``surface conditions'', ``how-to-paint'' of item ``Moon''.

\noindent $\bullet$ \textbf{ \textit{Videos}} is a set of short videos whose duration may not exceed 5 minutes. In this paper, we only focus on knowledge videos we detected, Where we follow OECD ~\citep{knowledgeoecd} to define knowledge as:
\begin{enumerate}
    \item \textbf{Know-what} refers to knowledge about \textit{facts}. E.g. How many people live in New York? 
    \item \textbf{Know-why} refers to scientific knowledge of the principles and laws of nature. E.g. Why does the earth revolve around the sun?
    \item \textbf{Know-how} refers to skills or the capability to do something. E.g. How to cook bacon in the oven.
\end{enumerate}
When the algorithm can extract the item-aspect pair of one video, this video can be \textit{linked} to Kuaipedia.


\section{Extraction Process}

\subsection{System Overview}
\begin{figure}[t]
\centering
\includegraphics[height=0.15\textwidth]{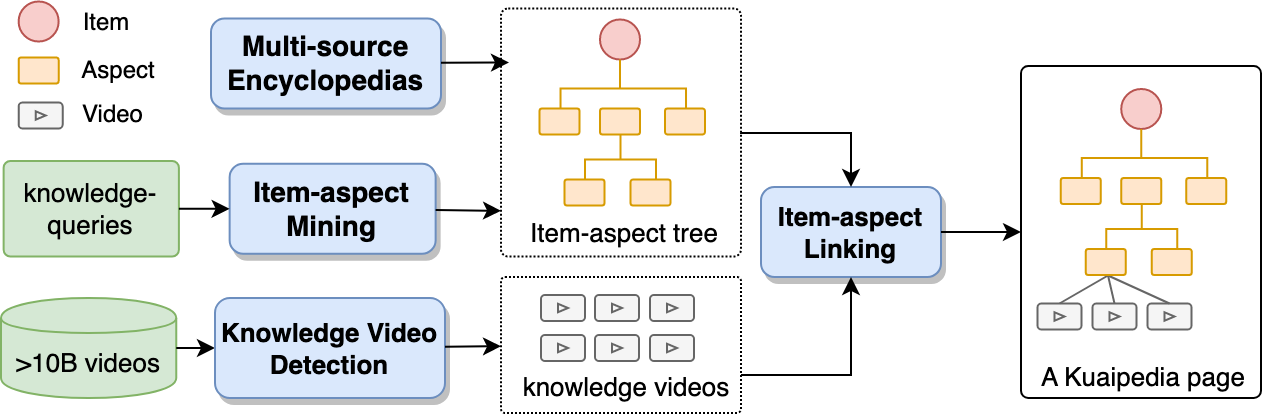}
\caption{Kuaipedia construction framework}\label{fig:framework}
\end{figure}

Our Kuaipedia construction process is outlined in Figure \ref{fig:framework}. To begin, we identify and extract knowledge videos from a vast number of videos. We then gather items from a variety of sources, e.g. Wikipedia, and mine aspects from knowledge-intensive queries to construct item-aspect trees. Finally, we utilize ``multi-modal item-aspect linking'' to associate these knowledge videos with the relevant item-aspect pairs.

\begin{figure*}[t]
\centering
\includegraphics[height=0.165\textwidth]{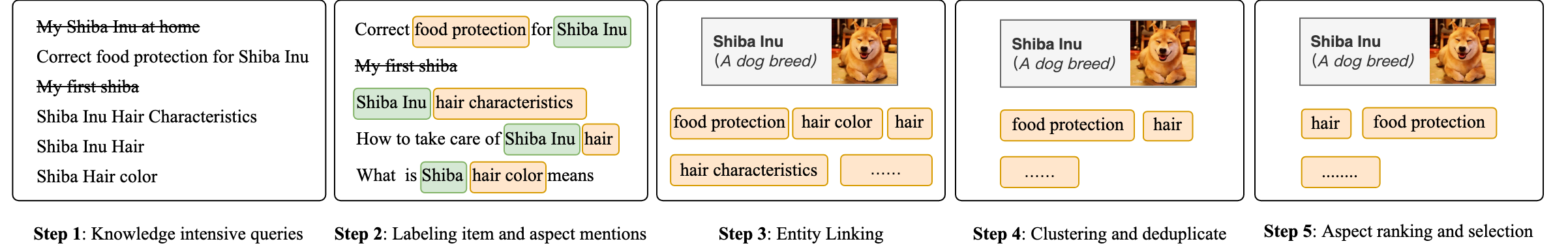}
\caption{A live example of the procedure of aspect mining.}\label{fig:aspect-mining}
\end{figure*}

\begin{figure*}[t]
\centering
\includegraphics[height=0.48\textwidth]{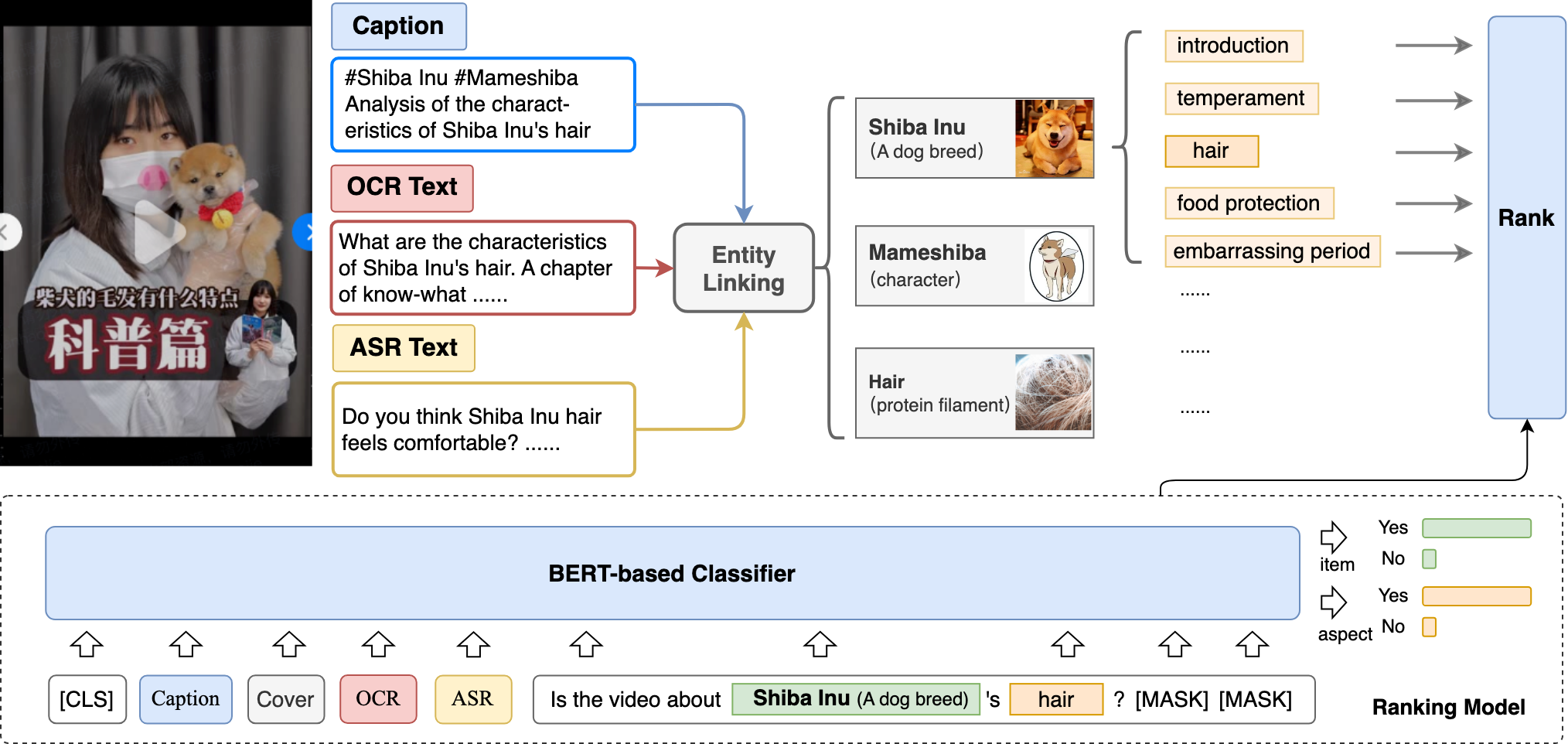}
\caption{Two phase Multi-modal Aspect Linking Model.}\label{fig:mmal}
\end{figure*}

\subsection{Knowledge Video Detection}
The initial step in creating Kuaipedia involves the selection of high-quality and high-knowledge-density videos, which form the foundation for all subsequent procedures. The task at hand involves determining whether a given short video $V = \{T_c, I_c, T_o, T_a\}$ is a knowledge video. The video is comprised of a user-edited caption $T_c$, the cover image $I_c$, the result of optical character recognition (OCR) applied to the video frames $T_o$, the result of automatic speech recognition (ASR) of the audio $T_a$.
We construct an multi-modal input of \texttt{\{[CLS],[MASK],$T_c$,[SEP],$T_o$,[SEP],$T_a$,[SEP]\}} , and using BERT~\citep{conf/naacl/DevlinCLT19} to encode, where the word embedding of the first \texttt{[MASK]} will be replaced by the result embedding of $ResNet(I_c)$~\citep{DBLP:journals/corr/HeZRS15}. A binary classifier was then trained to categorize a video into a knowledge video or not.

\subsection{Item-aspect Mining}
\label{Section:Item-aspect Mining}
As previously established, items consist of both entities and concepts, while aspects are keywords or keyphrases associated with these items. The challenge lies in extracting the "useful" and "user-centered" aspects relevant to a specific item, which can be obtained from sources such as Wikipedia. To address this challenge, we have devised a five-step mining process, which is outlined as follows.

First, we selected to extract aspects from "knowledge-intensive queries" that meet the following criteria:
\begin{enumerate}
\item The top-ranked video of the query must have received more than 5 clicks and a click rate greater than 80\%.
\item More than half of the clicked videos of the query must be classified as knowledge videos.
\end{enumerate}
Second, for each query, we first identified both the item and aspect mentions. We trained a team of annotators to distinguish between item and aspect spans and then utilized a sequence labeling task to extract these mentions. The task was accomplished using a combination of BERT, biLSTM, and CRF algorithms\footnote{\url{https://github.com/macanv/BERT-BiLSTM-CRF-NER}}.
Third, after extracting the item spans, we trained a BERT-based entity disambiguation module ~\citep{chunguangel} to link the item mentions to our existing encyclopedia.
Fourth, we grouped queries with the same items and generated an embedding for each query. We then applied a clustering algorithm to detect duplicates and selected the query closest to the cluster center as the master query. The aspect extracted from the master query became the master aspect of the cluster.
Fifth, for each item, we ranked the master aspects based on the aggregated search views and selected the top 100 as the final mined results. A relation $R_1$ was established between the item and the mined aspect. We also manually assigned parent aspects to some typical aspects to build a hyponymy-and-hypernymy relation $R_2$ between aspects.
Finally, we constructed the Item-aspect trees (IAT) $\HM'$ through these five steps, as demonstrated in Figure \ref{fig:aspect-mining}.

\subsection{Multi-modal Item-aspect Linking}
\label{Section:Multi-modal Item-aspect Linking}
After establishing the relationships between items and aspects, the next phase involves linking the knowledge videos to each item-aspect pair. To achieve this, we introduce a new task named "Multi-modal Item-aspect Linking": given a knowledge video $V$ and the Item-Aspect Trees(IAT) $\HM'$, our objective is to associate $V$ with the most suitable item-aspect pair $ \{ I, R_1, A \} \in \HM'$. To solve this problem, we propose a two-phase Multi-modal Aspect Linking Model (MMAL), which is illustrated in Figure \ref{fig:mmal}.

To begin with, we require a \textbf{recall module} that can recognize all the items mentioned in a knowledge video $V$ along with its caption, OCR text, and ASR text. The module then links the mentions to the corresponding entries in our encyclopedia through the use of Entity Linking techniques such as those described in ~\citep{chunguangel}.

Then we need a \textbf{ranking module}. Given a list of items $\mathbf{I} = \{ I_1, I_2, ..., I_N \}$, where $N$ is the number of candidate items. For each item $I_i$, we gather all the related aspects and flatten them into a list $\mathbf{A}_{i} = \{ A_{i, 1}, A_{i, 2}, ..., A_{i, k_i}\}$, where $k_i$ represents the total number of aspects associated with $I_i$. This results in $K = \sum_{i = 1}^{N}k_i$ item-aspect pairs ${\{ (I_i, A_{i, j})\}_{i=1}^{N}}_{j=1}^{k_i}$. Subsequently, we train a binary classifier to evaluate the relevance between the video and each item-aspect pair.

We generate the video context by combining the user-edit caption $T_c$, Cover Image $I_c$, OCR Text $T_o$, and ASR Text $T_a$ for each video frame. To enhance the understanding of our task by pre-trained language models, we develop a task-specific prompt input. The context and template are structured as follows:

\par \textbf{Context}: \texttt{[CLS]} \texttt{[MASK]} Caption \texttt{[SEP]} OCR \texttt{[SEP]} ASR \texttt{[SEP]}
\par \textbf{Prompt}: Is the video about \texttt{Item-title} (\texttt{Item-subtitle}) 's \texttt{Aspect-name} ? \texttt{[MASK]} \texttt{[MASK]}

Where \texttt{Item-title} is the title of the item wiki page (E.g. Shiba Inu) and \texttt{Item-subtitle} is the subtitle of this Item (E.g. A dog breed). \texttt{Aspect-name} is the text of the Aspect (E.g. Hair). The first \texttt{[MASK]} in context is a placeholder and the embedding of Transformer in this position will be replaced by by the result embedding of $ResNet(I_c)$, while the last two \texttt{[MASK]}s are the placeholder for the predicting word ``yes/no'' for the item and the aspect.

We applied a pre-trained language model such as BERT~\citep{conf/naacl/DevlinCLT19} to represent the context and prompt, and then used cross-entropy loss and stochastic gradient descent to optimize it. Finally we obtained scores for each item-aspect pair $(I_i, R_1, A_{i,j})$ relative to the video $V$ as $s_{i,j}$. The linked item-aspect pair $(I,A)$ of $V$ was determined as:

\begin{equation}
 (I, R_1, A) = argmax_{s_{i, j}} {\{ (I_i, R_1, A_{i, j})\}_{i=1}^{N}}_{j=1}^{k_i}, and ~ s_{i, j} > \theta
\end{equation}
Where $\theta$ is a predetermined threshold used to determine whether to keep the top pair or not.

\section{Intrinsic Evaluation}

\subsection{Item-Aspect Mining}
After the first step's filtering, we left 15 million queries as ``knowledge intention queries''. A sample of one thousand queries was taken, resulting in an accuracy of 90\%. And then we build a dataset of (\textit{query}, \textit{linked item}, \textit{aspect}) to train and evaluate the model performance in Step 2 and Step 3. The sequence labeling model in Step 2 achieves 80.4\% precision with respect to the \textit{item} and 67.4\% to the\textit{aspect}. After the entity disambiguation model in Step 3, the precision of \textit{linked item} drops to 77.6\%. 

\begin{figure}[t]
\centering
\includegraphics[width=\linewidth]{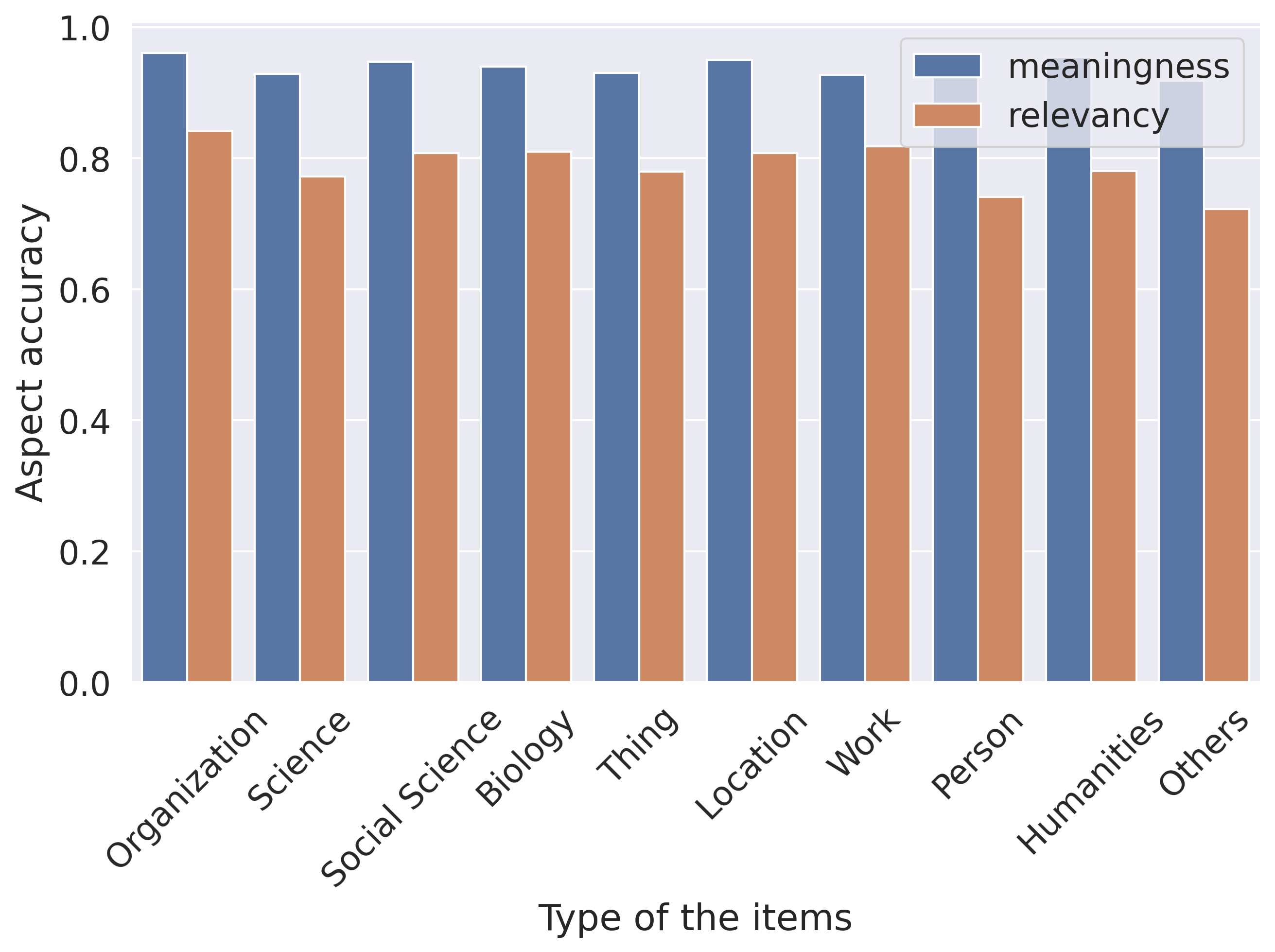}
\caption{Aspect accuracy distribution via the types of items}\label{fig:aspect-acc}
\end{figure}

We evaluated mined aspects using a criterion assessing their "meaningfulness" and "relevancy." To be meaningful, an aspect must be a valid word/phrase, have independent semantic meanings, and not be overly specific. To be relevant, it must be related to the item in common sense or make sense when searched online.
A sample of 10k item-aspect pairs was taken, and 5 human annotators evaluated each pair. The agreement between annotators was measured using the Kappa metric ($\kappa$) ~\citep{McHugh2012}. The 5 annotators achieved $\kappa_1 = 0.92$ for meaningfulness and $\kappa_2 = 0.97$ for relevancy. Table \ref{table:asp-exp} shows the evaluation results of the mined aspects, revealing that they are highly accurate with 91.1\% being meaningful and 77.1\% being both meaningful and relevant to the item. Figure \ref{fig:aspect-acc} displays the aspect accuracy distribution across different types of items, showing that certain types, such as "organization" or "location", may have higher relevancy, while others, such as "person", may not.

\begin{table}[t] 
 \centering
 \begin{tabular}{lc} 
  \toprule
  Item & value \\
  \midrule
   \#sampled item-aspect pairs for evaluation & 10,000 \\
   accuracy (meaningful) & 91.1\% \\
   accuracy (meaningful + relevant) & 77.1\% \\
 \bottomrule 
 \end{tabular} 
 
 \caption{
    Human evaluation results of the item-aspect mining.
 } 
\label{table:asp-exp}
\end{table}

\begin{figure}[t]
\centering
\includegraphics[width=\linewidth]{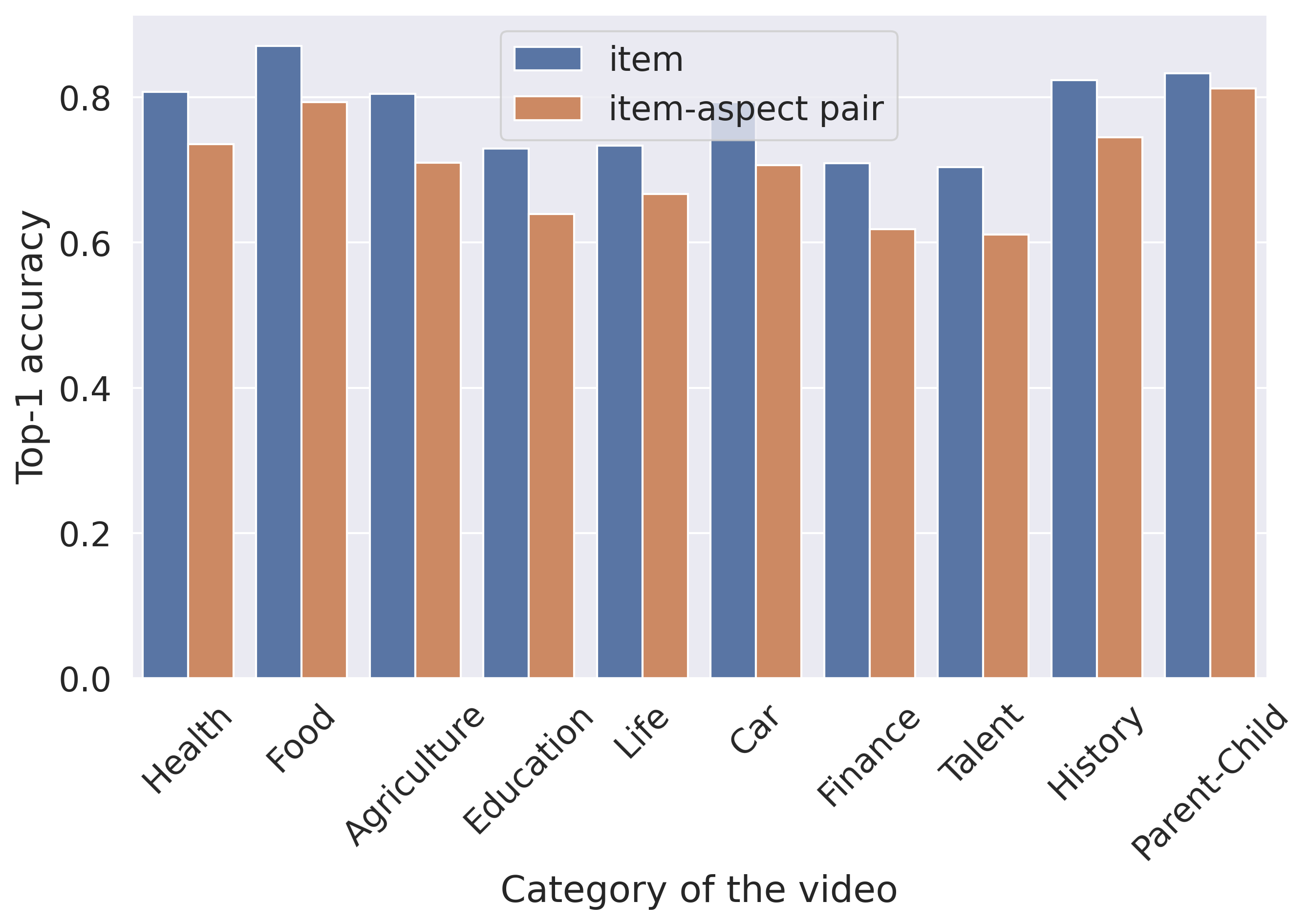}
\caption{Item-aspect linking accuracy distribution via the categories of videos (Here we only select top 10 categories)}\label{fig:linking-acc}
\end{figure}

\subsection{Multi-modal Item-aspect Linking}
\noindent \textbf{Dataset} As for text entity linking in the recall module, we use CCKS19 dataset ~\citep{sun-etal-2022-nsp} to train the overall module. For the ranking module,  more than 40k (\textit{video}, \textit{item}, \textit{aspect}, \textit{label}) quadruples were collected through training by a group of annotators, where the label consists of a pair of words, each being either "yes" or "no," indicating whether the video is related to the aspect of the item.

In order to evaluate the correlation between an item-aspect pair and a short video, we've devised a a detailed four-tiered criteria (Exactly Relevant, Moderately Relevant, Slightly Relevant, and Irrelevant). Annotators first determine item relevance, considering factors such as topic alignment and scope, then move onto aspect relevance, which looks at content coverage and semantic congruence.

The inter-annotator agreement, measured by Kappa, was $0.88$ among the 5 annotators hired for this dataset. We also split it into training set and testing set, whose size is 32,962 and 10,155 separately.


\noindent \textbf{Metric}: To evaluate the recall module, we select only the ``yes'' labeled examples and measure the recall@N for both items and aspects, intending to find the ground truth pair without regard for their ranking. The rank module, on the other hand, is evaluated using precision and recall metrics for both items and aspects.

\noindent \textbf{Experimental Setting}:We have implemented several baseline models for comparison, including: (1) Random: This model predicts yes/no outcomes based on random guessing. (2) LR: In this approach, we concatenate the embeddings of the cover image and the item-aspect text sentence to form input features, and then employ logistic regression for classification. (3) T5-small and BERT-base: We utilize T5-small (77M) and BERT-base (110M) as the backbone encoders, feeding the context described in Section~\ref{Section:Multi-modal Item-aspect Linking} as input, followed by a linear layer for classification. (4) GPT-3.5: We manually designed a template prompt for GPT-3.5-turbo, enabling it to generate classification results in a zero-shot setting. For our model, we use BERT-base as our backbone encoder and the learning rate is 1e-5, the number of epochs is 3 and the batch size is 32.

 

\begin{table}[!t] 
 \centering
 \begin{tabular}{l|cccc} 
  \toprule
  Model & Item P & Item R & Item-aspect P  & Item-aspect R \\
  \midrule
  Random & 87.7\% & 49.8\% & 36.4\% & 49.6 \% \\
  LR & 90.4\% & 68.3\% & 55.1\% & 2.7 \% \\
  T5-small & 93.7\% & 76.1\% & 79.3\% & 58.5 \% \\
  BERT-base & 94.3\% & 77.8\% & 81.5\% & 62.7 \% \\
  GPT-3.5 & 90.5\% & 86.4\% & 41.8\% & 95.7 \% \\
  \midrule
  Ours &  94.7\% & 79.7\% & 83.0\% & 65.7 \% \\
 \bottomrule 
 
 \end{tabular} 
 \caption{
    Experimental results of the ranking module of multi-modal aspect linking. 
 } 
\label{table:asp-link-exp}
\end{table}

\begin{table}[!t] 
 \centering
 \begin{tabular}{lc} 
  \toprule
  Item & value \\
  
  
  
  \midrule
  \#items (CN-Wikipedia) & 1,256,000  \\
  \#items (CN-DBPedia) & 10,341,196  \\
  \#items (ours) & > 26 millions  \\
  \#aspects & > 2.5 millions \\
  \#videos & > 200 millions \\
  
  \midrule
  \#item-aspect pairs & ~70 millions \\
  \#item-aspect pairs (Have video linked) & $~$ 1 million \\
  \#item-aspect-video triplets & $~$ 100 millions \\
  
  \midrule
  \#item-aspect Top1 video accuracy (to item) & 90.0\% \\
  \#item-aspect Top1 video accuracy (to pair) & 82.8\% \\
  \bottomrule 
  
 \end{tabular} 
 \caption{
    Overall statistics of Kuaipedia.
 } 
\label{table:overall-stat}
\end{table}

\noindent \textbf{Experimental Results} The recall module of multi-modal aspect linking can identify ground items for nearly 94\% of the videos and also find 88\% of the ground aspects when the ground items have been identified. As shown in Table \ref{table:asp-link-exp}, after the application of the ranking module, our model achieve the best results and  attain a precision of 83.0\% with 65.7\% of the videos having true item-aspect pairs. There is a close precision between item and item-aspect pair, however, the recall differs by 12.3\%, which may be attributed to the absence of relevant aspects in Kuaipedia for the videos to link to. Figure \ref{fig:linking-acc} illustrates the distribution of linking accuracy across different categories of videos. It can be observed that linking videos to item-aspect pairs is easier in categories such as Food'', Parent-child'', and History'' due to their well-defined topics, whereas categories such as Finance'' and ``Talent'' may be more challenging.

\begin{figure*}[t]
\centering
\includegraphics[height=0.19\textwidth]{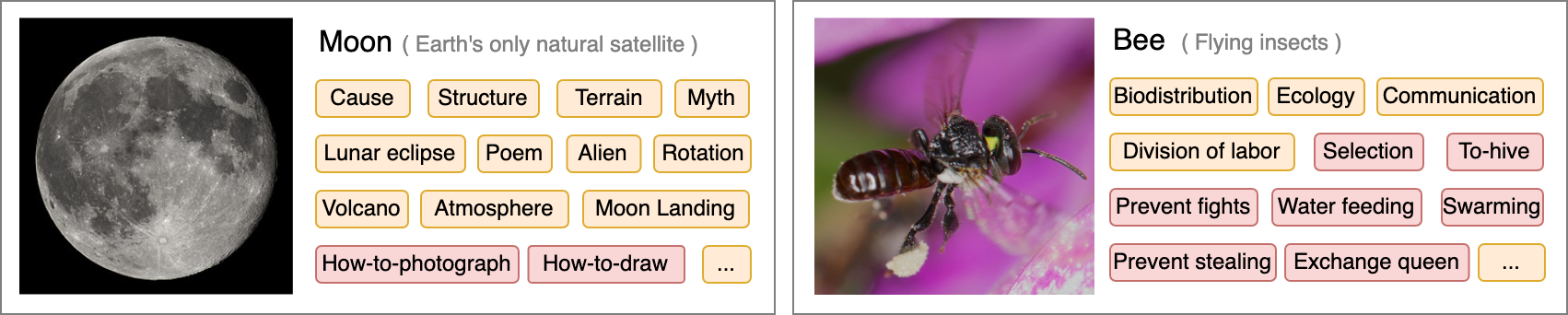}
\caption{Cases of mined aspects in Kuaipedia. The aspects in orange boxes are aspects that convey the knowledge of ``know-what'' or ``know-why'', while those in red boxes convey the knowledge of ``know-how''.}\label{fig:aspect-case}

\end{figure*}

\begin{figure*}[t]
\centering
\includegraphics[height=0.26\textwidth]{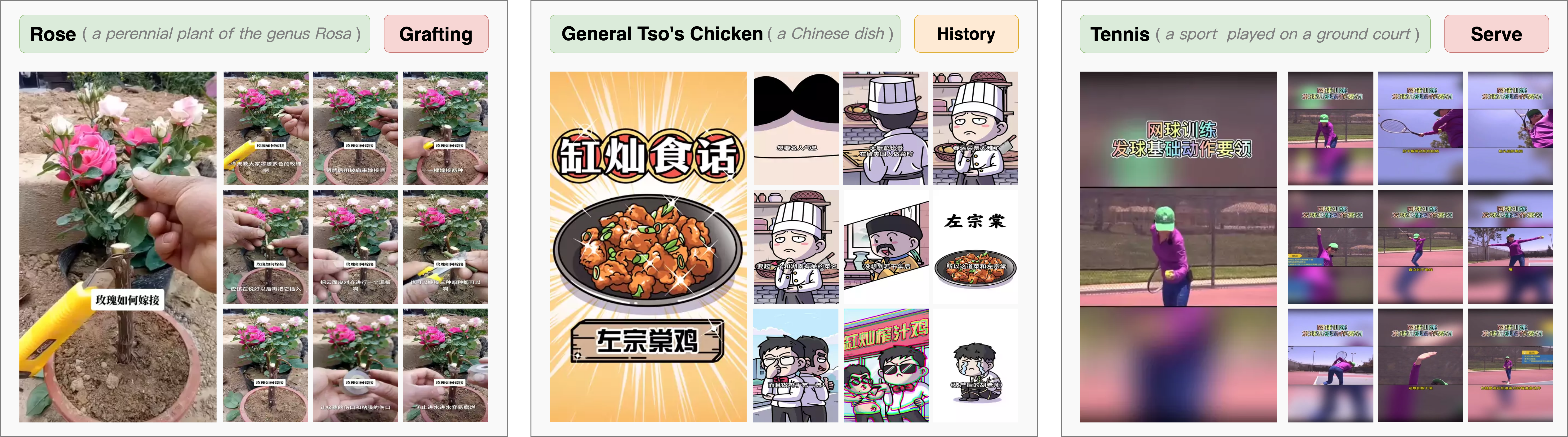}
\caption{Cases of linked videos to item-aspect pairs. Here we extract the covers and frames of the short videos. }\label{fig:video-case}
\end{figure*}

\subsection{Overall Statistics}
Here, we analyze Kuaipedia from three different perspectives, as depicted in Table \ref{table:overall-stat}.

\noindent \textbf{Node Scalability}: Kuaipedia demonstrates scalability, with a corpus that comprises over 25 million items and 2.5 million aspects (twice as many as CN-DBPedia~\citep{DBLP:conf/ieaaie/XuXLXLCX17} and 20 times more than CN-Wikipedia\footnote{As indicated by the number of articles on the website https://www.wikipedia.org/}). These aspects almost fully capture the meaning of entities and concepts in the Chinese language. The number of distinct aspects is reasonable and closely aligns with the usage of Chinese words or phrases. In addition, Kuaipedia features a vast collection of knowledge videos, numbering over 200 million in size.

\noindent \textbf{Edge Scalability}: We have extracted around 70 million edges between items and aspects, with nearly one million pairs linking to a knowledge video. There are 100 million videos that can be linked to item-aspect pairs, equating to 40\% - 50\% of the total. This disparity highlights that linked short videos are not evenly distributed, with many popular items and aspects having a higher number of videos to explain them.

\noindent \textbf{Quality}: To evaluate the relevancy of top-performing videos for item-aspect pairs, we employed and trained annotators. The annotators' report indicated that almost 90\% of the videos are relevant to the items and 82.8\% are relevant to the item-aspect pairs. This signifies that Kuaipedia is of high quality and accuracy.

\subsection{Case Study}
We analyze Kuaipedia from three perspectives, as depicted in Table \ref{table:overall-stat}. To showcase the richness of Kuaipedia, we present some intriguing examples in Figures \ref{fig:aspect-case} and \ref{fig:video-case}. Kuaipedia boasts millions of items that span a wide range of themes, including science, society, animals, health, finance, food, cars, sports, people, locations, organizations, and more. Figure \ref{fig:aspect-case} displays two items from distinct themes, demonstrating how the distribution of knowledge aspects, such as ``know-what'' ``know-why'' and ``know-how'' can vary. For instance, when discussing the ``Moon'' users tend to seek knowledge about its causes, structure, myths, and poems, whereas those discussing bees under the "animals" theme may focus on beekeeping and farming.
Figure \ref{fig:video-case} displays linked results of short videos to item-aspect pairs. It is evident that some ``know-how'' aspects, such as ``Grafting'' in the context of ``Rose'' and ``Serve'' in the context of ``Tennis'' are best taught through instructional videos. For "know-what" aspects, such as ``History'' authors may opt to create high-quality animations to convey the knowledge efficiently. One can refer to appendix \ref{more-pairs-case} for more insightful cases.


\section{Extrinsic Evaluation}
Given an item, we can enhance its representation by using corresponding aspects and videos from Kuaipedia. To verify the effectiveness of Kuaipedia, we conduct experiments on the CCKS19 dataset ~\citep{sun-etal-2022-nsp} for the common \textbf{Entity Typing} and \textbf{Entity Linking} tasks. We observe that items belonging to certain types, such as TVPlay or Human, in the knowledge base have limited knowledge or are rarely represented in short video platforms. Thus, we carefully select 18 types and exclude items of these types from our dataset. The detailed statistics of the final dataset are presented in Table \ref{table:statistics extrinsic evaluation}. Additionally, we demonstrate how Kuaipedia can be utilized to provide more informative prompts for large language models.


\subsection{Entity Typing}
\label{section:Entity Typing}
\noindent \textbf{Model}
Given a mention in a sentence, the aim of Entity Typing is to identify its types.
We design a baseline model for the task. Specifically, we use BERT as an encoder to get [CLS] representation and use MLP as a classifier to get logits of all types. Next, a threshold is set to identify which types are selected. We construct a vanilla input and an enhanced input for comparison as follows,

\par \textbf{Vanilla Input:} \texttt{[CLS]} \texttt{Mention} \texttt{Context} \texttt{[SEP]}
\par \textbf{Enhanced Input:} \texttt{[CLS]} \texttt{Mention} \texttt{Context} \texttt{[SEP]} \texttt{$A_1$} \texttt{[MASK]} \texttt{[SEP]} \texttt{$A_2$} \texttt{[MASK]} \texttt{[SEP]} \texttt{...}


\noindent Where \texttt{Mention} is in a given short text, i.e., \texttt{Context}. If \texttt{Mention} can be retrieved from Kuaipedia, we sample some aspects from items with the same name or synonym to construct the enhanced input, otherwise, the enhanced input degenerates into vanilla input. 
\texttt{$A_i$} is an aspect from $\mathcal{A}$ via Item-aspect Mining as described in \ref{Section:Item-aspect Mining}. The word embedding of \texttt{[MASK]} is replaced by video embedding $V_i$ of item-aspect pair $(I, A_i)$, where $V_i$ is from top 1 video via Multi-modal Item-aspect Linking as described in \ref{Section:Multi-modal Item-aspect Linking}. Overall, we use item-related aspects and video embeddings to enhance input.

\noindent \textbf{Experimental Results}
We compare vanilla input with enhanced input in terms of Precision, Recall, and F1 scores. The experimental results are reported in Table \ref{table:results extrinsic evaluation}. We observe that using enhanced input outperforms using vanilla input consistently under three metrics. It indicates that utilizing the aspects and videos from Kuaipedia can improve the performance of the model effectively.

\subsection{Entity Linking}
\noindent \textbf{Model}
Given a mention in a sentence, the aim of Entity Linking is to link it to a unique entity from a knowledge base. Similar to Entity Typing, we design different inputs for BERT to train a classifier. We construct positive and negative examples using all entities that have the same name or are synonymous with the mention. A binary classifier is trained to identify if the link is correct. The vanilla and enhanced inputs are constructed as follows,

\par \textbf{Vanilla Input:} \texttt{[CLS]} \texttt{Mention} \texttt{Context} \texttt{[SEP]} \texttt{Item} \texttt{Text} \texttt{[SEP]}
\par \textbf{Enhanced Input:} \texttt{[CLS]} \texttt{Mention} \texttt{Context} \texttt{[SEP]} \texttt{Item} \texttt{Text} \texttt{[SEP]} \texttt{$A_1$} \texttt{[MASK]} \texttt{[SEP]} \texttt{$A_2$} \texttt{[MASK]} \texttt{[SEP]} \texttt{...}

\noindent Where \texttt{Text} is the description about entity \texttt{Item} in the knowledge base and we use all aspects of \texttt{Item} to enhance the input. The meanings of other symbols are the same as those described in \ref{section:Entity Typing}.

\noindent \textbf{Experimental Results}
Similar to Entity Typing, we evaluate Precision, Recall, and F1 for vanilla and enhanced inputs and the results are reported in Table \ref{table:results extrinsic evaluation}. We observe that using enhanced input can improve the performance of the model under F1 metric. This improvement is attributed that the aspects and videos from Kuaipedia can effectively enhance the entity information and help the model get a more accurate judgment.

\begin{table}[!t]
\centering
\begin{tabular}{llc}
\toprule
Type                            & Item                & Value \\ \midrule
\multirow{4}{*}{Knowledge Base} & \#item              & 34528 \\
                                & \#type             & 33    \\
                                & \#item (w/ aspects) & 14570 \\
                                & \#item-aspect pair  & 108392\\ \midrule
\multirow{2}{*}{Train Dataset}  & \#sentence          & 68367      \\
                                & \#mention           & 136091      \\ \midrule
\multirow{2}{*}{Dev Dataset}    & \#sentence          & 7611      \\
                                & \#mention           & 15059      \\ \bottomrule
\end{tabular}
\caption{
    Statistics of the knowledge base and dataset for extrinsic evaluation. Since the test set is not public, we use the dev set to test our model.
}
\label{table:statistics extrinsic evaluation}
\end{table}

\begin{table}[!t]
\centering
\begin{tabular}{llccc}
\toprule
Task                            & Input    & P             & R              & F1             \\ \midrule
\multirow{2}{*}{Entity Typing}  & Vanilla  & 97.79         & 97.07          & 97.43          \\
                                & Enhanced & \textbf{98.8} & \textbf{97.44} & \textbf{98.12} \\ \midrule
\multirow{2}{*}{Entity Linking} & Vanilla  & 74.82         & \textbf{83.77}          & 79.04          \\
                                & Enhanced & \textbf{78.91}& 83.67          & \textbf{81.22} \\ \bottomrule
\end{tabular}
\caption{
    Comparison of model experiment results for the Entity Typing and Entity Linking task.
}
\label{table:results extrinsic evaluation}
\vspace{-5pt}
\end{table}


\begin{figure*}[t]
\centering 
\includegraphics[height=0.31\textwidth]{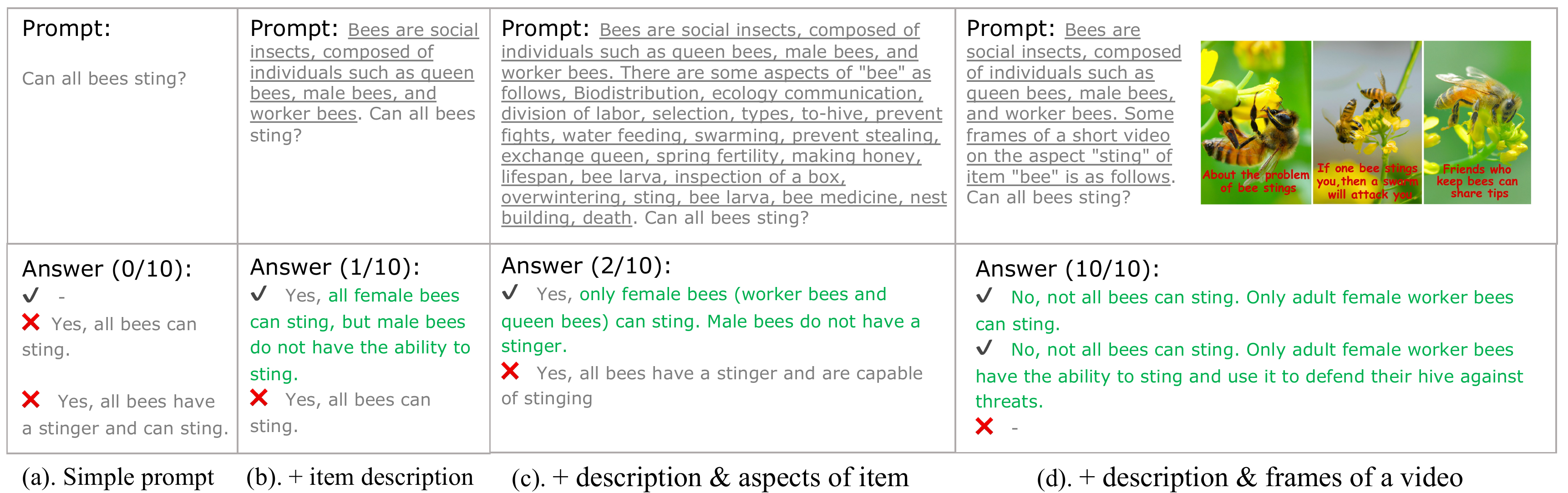}
\caption{Case of how Kuaipedia can help ChatGPT improve factuality capability. The text marked in blue is the correct reply content, while the rest are unreasonable. }\label{fig:chatgpt-case2}
\end{figure*}

\begin{figure*}[t]
\centering
\includegraphics[height=0.4\textwidth]{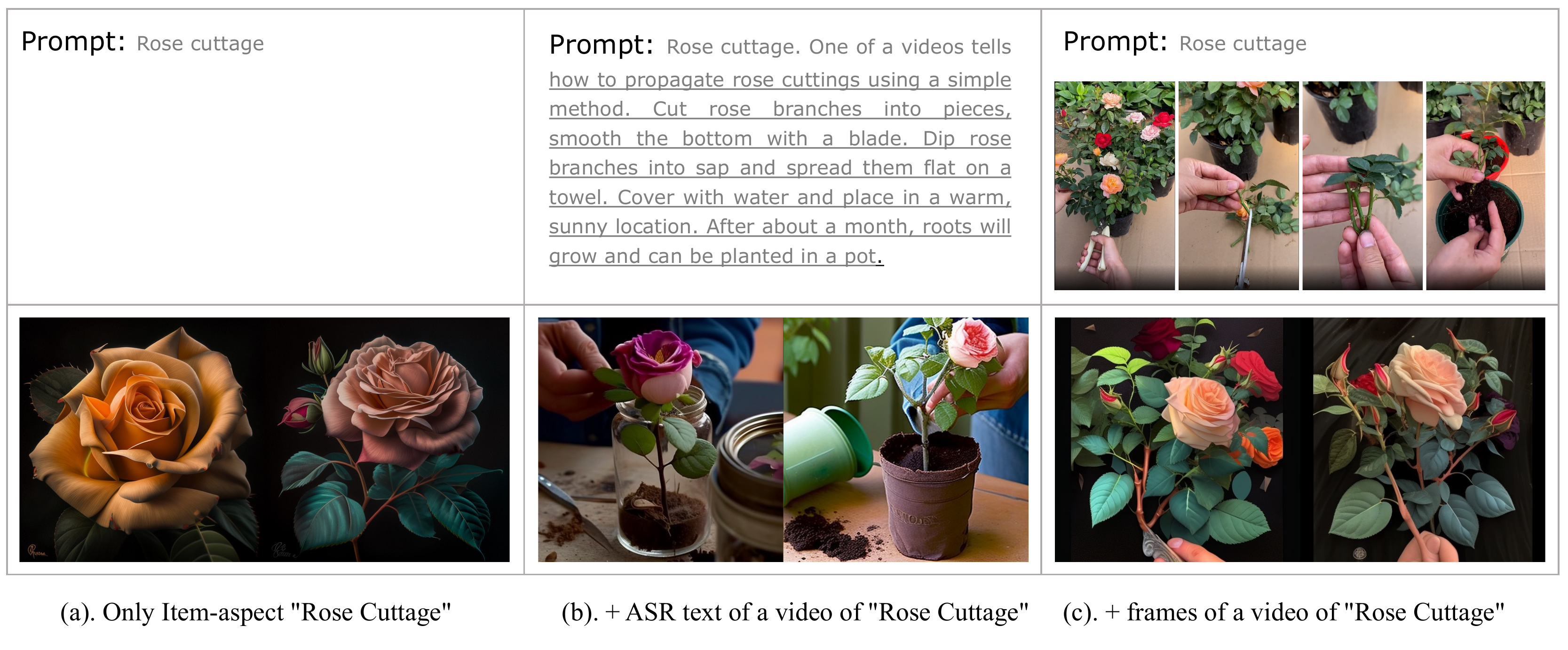}
\caption{Case of how Kuaipedia can help AI artisan ``imagine'' the action of ``Rose cuttage''. }\label{fig:video-case}
\end{figure*}

\subsection{Better Prompting}
Recently, the remarkable success of large language models such as GPT-3 \citep{Brown2020LanguageModels} and Dall·E\cite{DBLP:journals/corr/abs-2204-06125} has garnered widespread attention among researchers. These models are being actively pursued in a range of tasks, particularly in the areas of few-shot or zero-shot learning. An important aspect of leveraging these powerful pre-trained models is the selection of appropriate prompts. In this paper, we demonstrate how to effectively utilize Kuaipedia to obtain superior prompts, thus enhancing the quality of the generated results. Our experiments were conducted using the publicly available toolkits ChatGPT~\footnote{https://chat.openai.com/. We use Jan 9 version.} and MidJourney~\footnote{https://midjourney.com/. We use v4 release.}.

\noindent
\textbf{Improve factuality capability}
Despite its remarkable conversational abilities, ChatGPT still has limitations in terms of factuality and mathematical proficiency. As demonstrated in Figure \ref{fig:chatgpt-case2} (a), when asked "Can all bees sting?" ten times, ChatGPT consistently replied "Yes". However, this is not accurate as male bees do not have stingers and therefore cannot sting. The prompt can be improved by incorporating existing descriptions of the entity, as shown in Figure \ref{fig:chatgpt-case2} (b). Further incorporation of aspects about the item can lead to more diverse and potentially correct answers, as seen in Figure \ref{fig:chatgpt-case2} (c). The highest accuracy was achieved when the top-ranked video about the item-aspect "bees sting" was added to the input prompt along with several OCR frames, as demonstrated in Figure \ref{fig:chatgpt-case2} (d). This demonstrates that incorporating multimodal information from Kuaipedia can help improve the accuracy of ChatGPT's answers.

\begin{figure*}[h]
\centering
\includegraphics[width=1.0\textwidth]{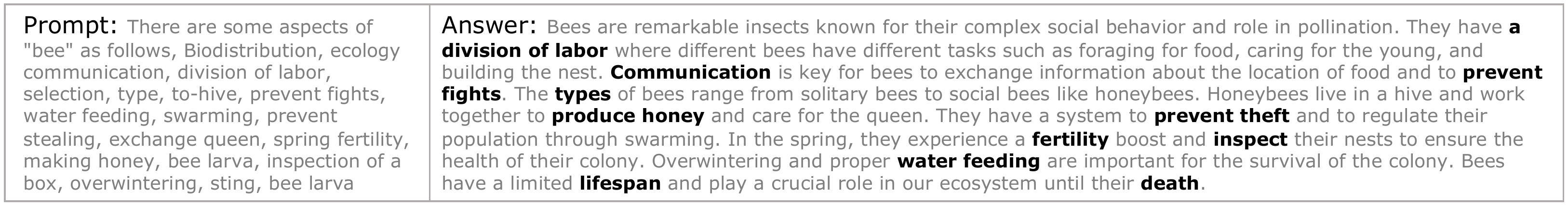}
\caption{Case of how Kuaipedia can help  ChatGPT generate more professional introductions. }\label{fig:chatgpt-case1}
\end{figure*}

\noindent
\textbf{Improve professionality}
In order to produce a comprehensive description of a certain item, we have employed a novel approach that combines the use of a simple prompt with relevant aspects from Kuaipedia. This approach enables ChatGPT to generate an introduction that is not only simple but also rich in detail. The example of ``a introduction for bees'' is shown in Figure \ref{fig:chatgpt-case1}

\noindent
\textbf{Generate images about the knowledge of "know-how"}
Text-to-image models like Dalle·E require appropriate prompts to generate the desired images. When it comes to knowledge of ``know-how'', however, users may struggle to provide comprehensive and accurate prompts, leading to subpar generation results. As seen in Figure \ref{fig:chatgpt-case2} (a), a simple prompt such as 'Rose Cuttage' only results in the generation of a rose, lacking the cuttage process. By adding some frames to the prompt, as shown in Figure \ref{fig:chatgpt-case2} (c), some cuttage actions are depicted. The results are improved further when the ASR text of the corresponding video of the item-aspect "Rose Cuttage" in Kuaipedia is added, as seen in Figure \ref{fig:chatgpt-case2} (b). The use of Kuaipedia allows the AI artisan to better ``imagine'' the action of ``rose cuttage'' with human hands, resulting in improved images that accurately depict the cuttage process. This highlights the benefit of using Kuaipedia in enhancing the generation of images related to the knowledge of "know-how".

\section{Related Work}

\subsection{Multimodal Knowledge Graph}
A multimodal knowledge graph (MMKG) is a graph-based representation of entities and their relationships, where each entity is described by multiple modalities, such as text, images, and videos. The ability to represent and reason about multiple modalities enables MMKGs to capture rich and diverse information and to support various applications, such as question answering, recommendation, and few-shot learning.

There has been a significant amount of research on building and maintaining MMKGs. NEIL~\citep{chen2013neil} uses a semi-supervised learning algorithm that jointly discovers common sense relationships and labels instances of the given visual categories. GIGA~\citep{DBLP:conf/acl/LiZLPWCWJCVNF20} is a structured knowledge base from heterogeneous multimedia data and enables the seamless search of complex graph queries and retrieves multimedia evidence including text, images, and videos. IMGpedia ~\citep{DBLP:conf/semweb/FerradaBH17} is a large-scale linked dataset that incorporates visual information of the images from the WIKIMEDIA COMMONS dataset. As well as IMGpedia, Image Graph~\citep{onoro-rubio2018answering}, MMKG~\citep{DBLP:journals/corr/abs-1903-05485}, Richpedia~\citep{DBLP:conf/jist/WangQWZ19}, VisualSem~\citep{DBLP:journals/corr/abs-2008-09150} are those MMKGs constructed by symbol grounding, which grounding visual content into symbols in existing KGs such as DBpedia~\citep{journals/semweb/LehmannIJJKMHMK15}.
There also has been a growing interest in using MKGs for various applications, such as question answering, recommendation, and few-shot learning. In ~\citep{Ding2022mukea} authors extract and accumulate multimodal knowledge for knowledge-based visual question answering. ~\citep{sun2020multi} incorporates multi-modal knowledge graph into recommender systems. In ~\citep{wang2020large}, the authors improve the performance of few-shot learning by utilizing both visual and textual information to find discriminative parts of objects.

Kuaipedia enriches the graph representations by grounding the short videos to entities or concepts in existing encyclopedias, rather than relying solely on images.

\subsection{Knowledge Extraction}
Here, we focus on two important tasks in knowledge extraction: Name entity recognition (NER) and entity linking (EL), which involve identifying and linking mentions of entities in the text to their corresponding entries in a knowledge base. These techniques play a crucial role in many natural language processing and information retrieval applications.
Numerous studies have been conducted in NER in recent years, aimed at improving the accuracy and scalability of these techniques. Traditional NER methods rely on hand-crafted rules, which can be designed based on domain-specific gazetteers~\citep{journals/ai/EtzioniCDPSSWY05,conf/lrec/SekineN04} and syntactic-lexical patterns~\citep{journals/jbi/ZhangE13}, while recent approaches utilize deep learning models such as recurrent neural networks (RNNs)~\citep{huang2015bidirectional,lample2016neural,chiu2015named,journals/corr/NguyenSDF16,conf/acl/ZhengWBHZX17} and convolutional neural networks (CNNs)~\citep{conf/medinfo/WuJLX15,conf/cncl/ZhouZXQBX17,conf/emnlp/StrubellVBM17} to capture contextual information in the input text. EL, on the other hand, is the task of linking mentions of entities to their unique identifiers in a knowledge base. One of the major challenges in EL is disambiguation, i.e., resolving the correct entity for an ambiguous mention. Several methods have been proposed to address this issue, including relation-based methods~\citep{kolitsas2018end,Yang2019,Le2019Boosting,Le2018}, or contextual BERT-driven language-based methods~\citep{Broscheit2019,DeCao2020AutoregressiveER,Yamada2020}.
Recently, there has been an increasing interest in incorporating other sources of information, such as images and videos, into NER and EL systems. Multi-modal NER and EL models ~\citep{moon-etal-2018-multimodal,yu-etal-2020-improving-multimodal,DBLP:conf/mm/GanLWWHH21,DBLP:conf/ecir/AdjaliBFBG20} have shown promising results in improving the performance of these systems, especially in low-resource scenarios. 

Here in Kuaipedia, we use the progress of multimodal NER and entity linking to improve our system.

\subsection{Pretrain Language Models}
Recently, the rapid emergence of large-scale pre-trained language models has brought the research frontiers of NLP to a new era~\cite{DBLP:journals/corr/abs-2003-08271}. 
Among these models, BERT~\cite{DBLP:conf/naacl/DevlinCLT19} is probably the most influential and popular model, which learns contextual token representations by a stack of transformer encoders, using two self-supervised learning objectives: masked language modeling and next sentence prediction. In this paper, we incorporate BERT as a basic encoder to do several tasks. Other the other head, the recent success of large-scale language models has led to growing interest in improving their capability to perform tasks via prompting~\citep{Brown2020LanguageModels}, which also boosts the generation capabilities of pre-train language models. In this paper, we research how Kuaipedia can help make better prompts.
\section{Conclusion}
In this paper, we introduce Kuaipedia, a large-scale multi-modal short-video encyclopedia. We first detect knowledge videos from more than ten billions short videos. And then we collect items from multi-sources and extract item-aspect pairs from user generated search queries. When the item-aspect trees are built, we apply ``multi-modal item-aspect linking'' techniques as a expansion of traditional ``entity linking'' to link short videos to item-aspect pairs. Kuaipedia is the first structured large-scale short-video encyclopedia that is organized by items, aspects, short videos and their relations. Both human and extrinsic evaluations prove that Kuaipedia is an effective and high-qulaity multi-modal short-video encyclopedia that has great potential in many downstream tasks and real-world applications.

\bibliographystyle{ACM-Reference-Format}
\bibliography{sample-base}


\begin{thebibliography}{56}


\ifx \showCODEN    \undefined \def \showCODEN     #1{\unskip}     \fi
\ifx \showDOI      \undefined \def \showDOI       #1{#1}\fi
\ifx \showISBNx    \undefined \def \showISBNx     #1{\unskip}     \fi
\ifx \showISBNxiii \undefined \def \showISBNxiii  #1{\unskip}     \fi
\ifx \showISSN     \undefined \def \showISSN      #1{\unskip}     \fi
\ifx \showLCCN     \undefined \def \showLCCN      #1{\unskip}     \fi
\ifx \shownote     \undefined \def \shownote      #1{#1}          \fi
\ifx \showarticletitle \undefined \def \showarticletitle #1{#1}   \fi
\ifx \showURL      \undefined \def \showURL       {\relax}        \fi
\providecommand\bibfield[2]{#2}
\providecommand\bibinfo[2]{#2}
\providecommand\natexlab[1]{#1}
\providecommand\showeprint[2][]{arXiv:#2}

\bibitem[\protect\citeauthoryear{Adjali, Besan{\c{c}}on, Ferret, Borgne, and
  Grau}{Adjali et~al\mbox{.}}{2020}]%
        {DBLP:conf/ecir/AdjaliBFBG20}
\bibfield{author}{\bibinfo{person}{Omar Adjali}, \bibinfo{person}{Romaric
  Besan{\c{c}}on}, \bibinfo{person}{Olivier Ferret},
  \bibinfo{person}{Herv{\'{e}}~Le Borgne}, {and} \bibinfo{person}{Brigitte
  Grau}.} \bibinfo{year}{2020}\natexlab{}.
\newblock \showarticletitle{Multimodal Entity Linking for Tweets}. In
  \bibinfo{booktitle}{\emph{Advances in Information Retrieval - 42nd European
  Conference on {IR} Research, {ECIR} 2020, Lisbon, Portugal, April 14-17,
  2020, Proceedings, Part {I}}} \emph{(\bibinfo{series}{Lecture Notes in
  Computer Science}, Vol.~\bibinfo{volume}{12035})}.
  \bibinfo{publisher}{Springer}, \bibinfo{pages}{463--478}.
\newblock


\bibitem[\protect\citeauthoryear{Alberts, Huang, Deshpande, Liu, Cho, Vania,
  and Calixto}{Alberts et~al\mbox{.}}{2020}]%
        {DBLP:journals/corr/abs-2008-09150}
\bibfield{author}{\bibinfo{person}{Houda Alberts}, \bibinfo{person}{Teresa
  Huang}, \bibinfo{person}{Yash Deshpande}, \bibinfo{person}{Yibo Liu},
  \bibinfo{person}{Kyunghyun Cho}, \bibinfo{person}{Clara Vania}, {and}
  \bibinfo{person}{Iacer Calixto}.} \bibinfo{year}{2020}\natexlab{}.
\newblock \showarticletitle{VisualSem: a high-quality knowledge graph for
  vision and language}.
\newblock \bibinfo{journal}{\emph{CoRR}}  \bibinfo{volume}{abs/2008.09150}
  (\bibinfo{year}{2020}).
\newblock


\bibitem[\protect\citeauthoryear{Bollacker, Evans, Paritosh, Sturge, and
  Taylor}{Bollacker et~al\mbox{.}}{2008}]%
        {DBLP:conf/sigmod/BollackerEPST08}
\bibfield{author}{\bibinfo{person}{Kurt~D. Bollacker}, \bibinfo{person}{Colin
  Evans}, \bibinfo{person}{Praveen~K. Paritosh}, \bibinfo{person}{Tim Sturge},
  {and} \bibinfo{person}{Jamie Taylor}.} \bibinfo{year}{2008}\natexlab{}.
\newblock \showarticletitle{Freebase: a collaboratively created graph database
  for structuring human knowledge}. In \bibinfo{booktitle}{\emph{Proceedings of
  the {ACM} {SIGMOD} International Conference on Management of Data, {SIGMOD}
  2008, Vancouver, BC, Canada, June 10-12, 2008}},
  \bibfield{editor}{\bibinfo{person}{Jason~Tsong{-}Li Wang}} (Ed.).
  \bibinfo{publisher}{{ACM}}, \bibinfo{pages}{1247--1250}.
\newblock
\urldef\tempurl%
\url{https://doi.org/10.1145/1376616.1376746}
\showDOI{\tempurl}


\bibitem[\protect\citeauthoryear{Broscheit}{Broscheit}{2019}]%
        {Broscheit2019}
\bibfield{author}{\bibinfo{person}{Samuel Broscheit}.}
  \bibinfo{year}{2019}\natexlab{}.
\newblock \showarticletitle{Investigating Entity Knowledge in {BERT} with
  Simple Neural End-to-End Entity Linking}. In
  \bibinfo{booktitle}{\emph{Proceedings of the 23rd Conference on Computational
  Natural Language Learning (CoNLL 2019)}}. \bibinfo{pages}{677--685}.
\newblock


\bibitem[\protect\citeauthoryear{Brown, Mann, Ryder, Subbiah, Kaplan, Dhariwal,
  Neelakantan, Shyam, Sastry, Askell, Agarwal, Herbert-Voss, Krueger, Henighan,
  Child, Ramesh, Ziegler, Wu, Winter, Hesse, Chen, Sigler, Litwin, Gray, Chess,
  Clark, Berner, McCandlish, Radford, Sutskever, and Amodei}{Brown
  et~al\mbox{.}}{2020}]%
        {Brown2020LanguageModels}
\bibfield{author}{\bibinfo{person}{Tom Brown}, \bibinfo{person}{Benjamin Mann},
  \bibinfo{person}{Nick Ryder}, \bibinfo{person}{Melanie Subbiah},
  \bibinfo{person}{Jared~D Kaplan}, \bibinfo{person}{Prafulla Dhariwal},
  \bibinfo{person}{Arvind Neelakantan}, \bibinfo{person}{Pranav Shyam},
  \bibinfo{person}{Girish Sastry}, \bibinfo{person}{Amanda Askell},
  \bibinfo{person}{Sandhini Agarwal}, \bibinfo{person}{Ariel Herbert-Voss},
  \bibinfo{person}{Gretchen Krueger}, \bibinfo{person}{Tom Henighan},
  \bibinfo{person}{Rewon Child}, \bibinfo{person}{Aditya Ramesh},
  \bibinfo{person}{Daniel Ziegler}, \bibinfo{person}{Jeffrey Wu},
  \bibinfo{person}{Clemens Winter}, \bibinfo{person}{Chris Hesse},
  \bibinfo{person}{Mark Chen}, \bibinfo{person}{Eric Sigler},
  \bibinfo{person}{Mateusz Litwin}, \bibinfo{person}{Scott Gray},
  \bibinfo{person}{Benjamin Chess}, \bibinfo{person}{Jack Clark},
  \bibinfo{person}{Christopher Berner}, \bibinfo{person}{Sam McCandlish},
  \bibinfo{person}{Alec Radford}, \bibinfo{person}{Ilya Sutskever}, {and}
  \bibinfo{person}{Dario Amodei}.} \bibinfo{year}{2020}\natexlab{}.
\newblock \showarticletitle{Language Models are Few-Shot Learners}. In
  \bibinfo{booktitle}{\emph{NeurIPS}}.
\newblock


\bibitem[\protect\citeauthoryear{Cao, Izacard, Riedel, and Petroni}{Cao
  et~al\mbox{.}}{2020}]%
        {DeCao2020AutoregressiveER}
\bibfield{author}{\bibinfo{person}{Nicola~De Cao}, \bibinfo{person}{Gautier
  Izacard}, \bibinfo{person}{Sebastian Riedel}, {and} \bibinfo{person}{Fabio
  Petroni}.} \bibinfo{year}{2020}\natexlab{}.
\newblock \showarticletitle{Autoregressive Entity Retrieval}. In
  \bibinfo{booktitle}{\emph{Proceedings of the 2021 International Conference on
  Learning Representations (ICLR 2021)}}.
\newblock


\bibitem[\protect\citeauthoryear{Chen, Shrivastava, and Gupta}{Chen
  et~al\mbox{.}}{2013}]%
        {DBLP:conf/iccv/ChenSG13}
\bibfield{author}{\bibinfo{person}{Xinlei Chen}, \bibinfo{person}{Abhinav
  Shrivastava}, {and} \bibinfo{person}{Abhinav Gupta}.}
  \bibinfo{year}{2013}\natexlab{}.
\newblock \showarticletitle{{NEIL:} Extracting Visual Knowledge from Web Data}.
  In \bibinfo{booktitle}{\emph{{IEEE} International Conference on Computer
  Vision, {ICCV} 2013, Sydney, Australia, December 1-8, 2013}}.
  \bibinfo{publisher}{{IEEE} Computer Society}, \bibinfo{pages}{1409--1416}.
\newblock
\urldef\tempurl%
\url{https://doi.org/10.1109/ICCV.2013.178}
\showDOI{\tempurl}


\bibitem[\protect\citeauthoryear{Chiu and Nichols}{Chiu and Nichols}{2015}]%
        {chiu2015named}
\bibfield{author}{\bibinfo{person}{Jason P.~C. Chiu} {and}
  \bibinfo{person}{Eric Nichols}.} \bibinfo{year}{2015}\natexlab{}.
\newblock \bibinfo{title}{Named Entity Recognition with Bidirectional
  LSTM-CNNs}.
\newblock
\newblock
\urldef\tempurl%
\url{http://arxiv.org/abs/1511.08308}
\showURL{%
\tempurl}
\newblock
\shownote{cite arxiv:1511.08308Comment: To appear in Transactions of the
  Association for Computational Linguistics.}


\bibitem[\protect\citeauthoryear{contributors}{contributors}{2006}]%
        {baidubaike}
\bibfield{author}{\bibinfo{person}{BaiduBaike contributors}.}
  \bibinfo{year}{2006}\natexlab{}.
\newblock \bibinfo{booktitle}{\emph{BaiduBaike}}.
\newblock
\urldef\tempurl%
\url{https://baike.baidu.com/}
\showURL{%
\tempurl}


\bibitem[\protect\citeauthoryear{contributors}{contributors}{1999}]%
        {investopedia}
\bibfield{author}{\bibinfo{person}{Investopedia contributors}.}
  \bibinfo{year}{1999}\natexlab{}.
\newblock \bibinfo{booktitle}{\emph{Investopedia}}.
\newblock
\urldef\tempurl%
\url{https://www.investopedia.com/}
\showURL{%
\tempurl}


\bibitem[\protect\citeauthoryear{contributors}{contributors}{2001}]%
        {wikipedia}
\bibfield{author}{\bibinfo{person}{Wikipedia contributors}.}
  \bibinfo{year}{2001}\natexlab{}.
\newblock \bibinfo{booktitle}{\emph{Wikipedia, The Free Encyclopedia}}.
\newblock
\urldef\tempurl%
\url{https://en.wikipedia.org}
\showURL{%
\tempurl}


\bibitem[\protect\citeauthoryear{Devlin, Chang, Lee, and Toutanova}{Devlin
  et~al\mbox{.}}{2019a}]%
        {DBLP:conf/naacl/DevlinCLT19}
\bibfield{author}{\bibinfo{person}{Jacob Devlin}, \bibinfo{person}{Ming{-}Wei
  Chang}, \bibinfo{person}{Kenton Lee}, {and} \bibinfo{person}{Kristina
  Toutanova}.} \bibinfo{year}{2019}\natexlab{a}.
\newblock \showarticletitle{{BERT:} Pre-training of Deep Bidirectional
  Transformers for Language Understanding}. In
  \bibinfo{booktitle}{\emph{NAACL-HLT}}. \bibinfo{pages}{4171--4186}.
\newblock


\bibitem[\protect\citeauthoryear{Devlin, Chang, Lee, and Toutanova}{Devlin
  et~al\mbox{.}}{2019b}]%
        {conf/naacl/DevlinCLT19}
\bibfield{author}{\bibinfo{person}{Jacob Devlin}, \bibinfo{person}{Ming-Wei
  Chang}, \bibinfo{person}{Kenton Lee}, {and} \bibinfo{person}{Kristina
  Toutanova}.} \bibinfo{year}{2019}\natexlab{b}.
\newblock \showarticletitle{BERT: Pre-training of Deep Bidirectional
  Transformers for Language Understanding}. In
  \bibinfo{booktitle}{\emph{NAACL-HLT}}. \bibinfo{pages}{4171--4186}.
\newblock


\bibitem[\protect\citeauthoryear{Ding, Yu, Liu, Hu, Cui, and Wug}{Ding
  et~al\mbox{.}}{2022}]%
        {Ding2022mukea}
\bibfield{author}{\bibinfo{person}{Yang Ding}, \bibinfo{person}{Jing Yu},
  \bibinfo{person}{Bang Liu}, \bibinfo{person}{Yue Hu},
  \bibinfo{person}{Mingxin Cui}, {and} \bibinfo{person}{Qi Wug}.}
  \bibinfo{year}{2022}\natexlab{}.
\newblock \showarticletitle{MuKEA: Multimodal Knowledge Extraction and
  Accumulation for Knowledge-based Visual Question Answering}. In
  \bibinfo{booktitle}{\emph{Proceedings of the IEEE/CVF Conference on Computer
  Vision and Pattern Recognition (CVPR)}}.
\newblock


\bibitem[\protect\citeauthoryear{Etzioni, Cafarella, Downey, Popescu, Shaked,
  Soderland, Weld, and Yates}{Etzioni et~al\mbox{.}}{2005}]%
        {journals/ai/EtzioniCDPSSWY05}
\bibfield{author}{\bibinfo{person}{Oren Etzioni}, \bibinfo{person}{Michael~J.
  Cafarella}, \bibinfo{person}{Doug Downey}, \bibinfo{person}{Ana-Maria
  Popescu}, \bibinfo{person}{Tal Shaked}, \bibinfo{person}{Stephen Soderland},
  \bibinfo{person}{Daniel~S. Weld}, {and} \bibinfo{person}{Alexander Yates}.}
  \bibinfo{year}{2005}\natexlab{}.
\newblock \showarticletitle{Unsupervised named-entity extraction from the Web:
  An experimental study.}
\newblock \bibinfo{journal}{\emph{Artif. Intell.}} \bibinfo{volume}{165},
  \bibinfo{number}{1} (\bibinfo{year}{2005}), \bibinfo{pages}{91--134}.
\newblock
\urldef\tempurl%
\url{http://dblp.uni-trier.de/db/journals/ai/ai165.html#EtzioniCDPSSWY05}
\showURL{%
\tempurl}


\bibitem[\protect\citeauthoryear{Ferrada, Bustos, and Hogan}{Ferrada
  et~al\mbox{.}}{2017}]%
        {DBLP:conf/semweb/FerradaBH17}
\bibfield{author}{\bibinfo{person}{Sebasti{\'{a}}n Ferrada},
  \bibinfo{person}{Benjamin Bustos}, {and} \bibinfo{person}{Aidan Hogan}.}
  \bibinfo{year}{2017}\natexlab{}.
\newblock \showarticletitle{IMGpedia: {A} Linked Dataset with Content-Based
  Analysis of Wikimedia Images}. In \bibinfo{booktitle}{\emph{The Semantic Web
  - {ISWC} 2017 - 16th International Semantic Web Conference, Vienna, Austria,
  October 21-25, 2017, Proceedings, Part {II}}} \emph{(\bibinfo{series}{Lecture
  Notes in Computer Science}, Vol.~\bibinfo{volume}{10588})},
  \bibfield{editor}{\bibinfo{person}{Claudia d'Amato}, \bibinfo{person}{Miriam
  Fern{\'{a}}ndez}, \bibinfo{person}{Valentina A.~M. Tamma},
  \bibinfo{person}{Freddy L{\'{e}}cu{\'{e}}}, \bibinfo{person}{Philippe
  Cudr{\'{e}}{-}Mauroux}, \bibinfo{person}{Juan~F. Sequeda},
  \bibinfo{person}{Christoph Lange}, {and} \bibinfo{person}{Jeff Heflin}}
  (Eds.). \bibinfo{publisher}{Springer}, \bibinfo{pages}{84--93}.
\newblock


\bibitem[\protect\citeauthoryear{Gan, Luo, Wang, Wang, He, and Huang}{Gan
  et~al\mbox{.}}{2021}]%
        {DBLP:conf/mm/GanLWWHH21}
\bibfield{author}{\bibinfo{person}{Jingru Gan}, \bibinfo{person}{Jinchang Luo},
  \bibinfo{person}{Haiwei Wang}, \bibinfo{person}{Shuhui Wang},
  \bibinfo{person}{Wei He}, {and} \bibinfo{person}{Qingming Huang}.}
  \bibinfo{year}{2021}\natexlab{}.
\newblock \showarticletitle{Multimodal Entity Linking: {A} New Dataset and {A}
  Baseline}. In \bibinfo{booktitle}{\emph{{MM} '21: {ACM} Multimedia
  Conference, Virtual Event, China, October 20 - 24, 2021}}.
  \bibinfo{publisher}{{ACM}}, \bibinfo{pages}{993--1001}.
\newblock


\bibitem[\protect\citeauthoryear{He, Zhang, Ren, and Sun}{He
  et~al\mbox{.}}{2015}]%
        {DBLP:journals/corr/HeZRS15}
\bibfield{author}{\bibinfo{person}{Kaiming He}, \bibinfo{person}{Xiangyu
  Zhang}, \bibinfo{person}{Shaoqing Ren}, {and} \bibinfo{person}{Jian Sun}.}
  \bibinfo{year}{2015}\natexlab{}.
\newblock \showarticletitle{Deep Residual Learning for Image Recognition}.
\newblock \bibinfo{journal}{\emph{CoRR}}  \bibinfo{volume}{abs/1512.03385}
  (\bibinfo{year}{2015}).
\newblock
\showeprint[arxiv]{1512.03385}
\urldef\tempurl%
\url{http://arxiv.org/abs/1512.03385}
\showURL{%
\tempurl}


\bibitem[\protect\citeauthoryear{Huang, Xu, and Yu}{Huang
  et~al\mbox{.}}{2015}]%
        {huang2015bidirectional}
\bibfield{author}{\bibinfo{person}{Zhiheng Huang}, \bibinfo{person}{Wei Xu},
  {and} \bibinfo{person}{Kai Yu}.} \bibinfo{year}{2015}\natexlab{}.
\newblock \bibinfo{title}{Bidirectional LSTM-CRF Models for Sequence Tagging}.
\newblock
\newblock
\urldef\tempurl%
\url{http://arxiv.org/abs/1508.01991}
\showURL{%
\tempurl}
\newblock
\shownote{cite arxiv:1508.01991.}


\bibitem[\protect\citeauthoryear{Kolitsas, Ganea, and Hofmann}{Kolitsas
  et~al\mbox{.}}{2018}]%
        {kolitsas2018end}
\bibfield{author}{\bibinfo{person}{Nikolaos Kolitsas},
  \bibinfo{person}{Octavian-Eugen Ganea}, {and} \bibinfo{person}{Thomas
  Hofmann}.} \bibinfo{year}{2018}\natexlab{}.
\newblock \showarticletitle{End-to-end neural entity linking}. In
  \bibinfo{booktitle}{\emph{Proceedings of the 22nd Conference on Computational
  Natural Language Learning (CoNLL 2018)}}. \bibinfo{pages}{519--529}.
\newblock


\bibitem[\protect\citeauthoryear{Lample, Ballesteros, Subramanian, Kawakami,
  and Dyer}{Lample et~al\mbox{.}}{2016}]%
        {lample2016neural}
\bibfield{author}{\bibinfo{person}{Guillaume Lample}, \bibinfo{person}{Miguel
  Ballesteros}, \bibinfo{person}{Sandeep Subramanian}, \bibinfo{person}{Kazuya
  Kawakami}, {and} \bibinfo{person}{Chris Dyer}.}
  \bibinfo{year}{2016}\natexlab{}.
\newblock \bibinfo{title}{Neural Architectures for Named Entity Recognition}.
\newblock
\newblock
\urldef\tempurl%
\url{http://arxiv.org/abs/1603.01360}
\showURL{%
\tempurl}
\newblock
\shownote{cite arxiv:1603.01360Comment: Proceedings of NAACL 2016.}


\bibitem[\protect\citeauthoryear{Le and Titov}{Le and Titov}{2018}]%
        {Le2018}
\bibfield{author}{\bibinfo{person}{Phong Le} {and} \bibinfo{person}{Ivan
  Titov}.} \bibinfo{year}{2018}\natexlab{}.
\newblock \showarticletitle{Improving entity linking by modeling latent
  relations between mentions}. In \bibinfo{booktitle}{\emph{Proceedings of the
  2018 Annual Meeting of the Association for Computational Linguistics (ACL
  2018)}}. \bibinfo{pages}{1595--1604}.
\newblock


\bibitem[\protect\citeauthoryear{Le and Titov}{Le and Titov}{2019}]%
        {Le2019Boosting}
\bibfield{author}{\bibinfo{person}{Phong Le} {and} \bibinfo{person}{Ivan
  Titov}.} \bibinfo{year}{2019}\natexlab{}.
\newblock \showarticletitle{Boosting entity linking performance by leveraging
  unlabeled documents}. In \bibinfo{booktitle}{\emph{Proceedings of the 2019
  Annual Meeting of the Association for Computational Linguistics (ACL 2019)}}.
  \bibinfo{pages}{1935--1945}.
\newblock


\bibitem[\protect\citeauthoryear{Lehmann, Isele, Jakob, Jentzsch, Kontokostas,
  Mendes, Hellmann, Morsey, van Kleef, Auer, and Bizer}{Lehmann
  et~al\mbox{.}}{2015}]%
        {journals/semweb/LehmannIJJKMHMK15}
\bibfield{author}{\bibinfo{person}{Jens Lehmann}, \bibinfo{person}{Robert
  Isele}, \bibinfo{person}{Max Jakob}, \bibinfo{person}{Anja Jentzsch},
  \bibinfo{person}{Dimitris Kontokostas}, \bibinfo{person}{Pablo~N. Mendes},
  \bibinfo{person}{Sebastian Hellmann}, \bibinfo{person}{Mohamed Morsey},
  \bibinfo{person}{Patrick van Kleef}, \bibinfo{person}{Sören Auer}, {and}
  \bibinfo{person}{Christian Bizer}.} \bibinfo{year}{2015}\natexlab{}.
\newblock \showarticletitle{DBpedia - A large-scale, multilingual knowledge
  base extracted from Wikipedia.}
\newblock \bibinfo{journal}{\emph{Semantic Web}} \bibinfo{volume}{6},
  \bibinfo{number}{2} (\bibinfo{year}{2015}), \bibinfo{pages}{167--195}.
\newblock
\urldef\tempurl%
\url{http://dblp.uni-trier.de/db/journals/semweb/semweb6.html#LehmannIJJKMHMK15}
\showURL{%
\tempurl}


\bibitem[\protect\citeauthoryear{Li, Zareian, Lin, Pan, Whitehead, Chen, Wu,
  Ji, Chang, Voss, Napierski, and Freedman}{Li et~al\mbox{.}}{2020}]%
        {DBLP:conf/acl/LiZLPWCWJCVNF20}
\bibfield{author}{\bibinfo{person}{Manling Li}, \bibinfo{person}{Alireza
  Zareian}, \bibinfo{person}{Ying Lin}, \bibinfo{person}{Xiaoman Pan},
  \bibinfo{person}{Spencer Whitehead}, \bibinfo{person}{Brian Chen},
  \bibinfo{person}{Bo Wu}, \bibinfo{person}{Heng Ji},
  \bibinfo{person}{Shih{-}Fu Chang}, \bibinfo{person}{Clare~R. Voss},
  \bibinfo{person}{Daniel Napierski}, {and} \bibinfo{person}{Marjorie
  Freedman}.} \bibinfo{year}{2020}\natexlab{}.
\newblock \showarticletitle{{GAIA:} {A} Fine-grained Multimedia Knowledge
  Extraction System}. In \bibinfo{booktitle}{\emph{ACL}}.
  \bibinfo{publisher}{Association for Computational Linguistics},
  \bibinfo{pages}{77--86}.
\newblock


\bibitem[\protect\citeauthoryear{Liu, Li, Garc{\'{\i}}a{-}Dur{\'{a}}n, Niepert,
  O{\~{n}}oro{-}Rubio, and Rosenblum}{Liu et~al\mbox{.}}{2019}]%
        {DBLP:journals/corr/abs-1903-05485}
\bibfield{author}{\bibinfo{person}{Ye Liu}, \bibinfo{person}{Hui Li},
  \bibinfo{person}{Alberto Garc{\'{\i}}a{-}Dur{\'{a}}n},
  \bibinfo{person}{Mathias Niepert}, \bibinfo{person}{Daniel
  O{\~{n}}oro{-}Rubio}, {and} \bibinfo{person}{David~S. Rosenblum}.}
  \bibinfo{year}{2019}\natexlab{}.
\newblock \showarticletitle{{MMKG:} Multi-Modal Knowledge Graphs}.
\newblock \bibinfo{journal}{\emph{CoRR}}  \bibinfo{volume}{abs/1903.05485}
  (\bibinfo{year}{2019}).
\newblock


\bibitem[\protect\citeauthoryear{McHugh}{McHugh}{2012}]%
        {McHugh2012}
\bibfield{author}{\bibinfo{person}{M.~L. McHugh}.}
  \bibinfo{year}{2012}\natexlab{}.
\newblock \showarticletitle{Interrater reliability: the kappa statistic}.
\newblock \bibinfo{journal}{\emph{Biochem Med (Zagreb)}} \bibinfo{volume}{22},
  \bibinfo{number}{3} (\bibinfo{year}{2012}), \bibinfo{pages}{276--82}.
\newblock


\bibitem[\protect\citeauthoryear{Moon, Neves, and Carvalho}{Moon
  et~al\mbox{.}}{2018}]%
        {moon-etal-2018-multimodal}
\bibfield{author}{\bibinfo{person}{Seungwhan Moon}, \bibinfo{person}{Leonardo
  Neves}, {and} \bibinfo{person}{Vitor Carvalho}.}
  \bibinfo{year}{2018}\natexlab{}.
\newblock \showarticletitle{Multimodal Named Entity Recognition for Short
  Social Media Posts}. In \bibinfo{booktitle}{\emph{Proceedings of the 2018
  Conference of the North {A}merican Chapter of the Association for
  Computational Linguistics: Human Language Technologies, Volume 1 (Long
  Papers)}}. \bibinfo{publisher}{Association for Computational Linguistics},
  \bibinfo{address}{New Orleans, Louisiana}, \bibinfo{pages}{852--860}.
\newblock
\urldef\tempurl%
\url{https://doi.org/10.18653/v1/N18-1078}
\showDOI{\tempurl}


\bibitem[\protect\citeauthoryear{Nguyen, Sil, Dinu, and Florian}{Nguyen
  et~al\mbox{.}}{2016}]%
        {journals/corr/NguyenSDF16}
\bibfield{author}{\bibinfo{person}{Thien~Huu Nguyen}, \bibinfo{person}{Avirup
  Sil}, \bibinfo{person}{Georgiana Dinu}, {and} \bibinfo{person}{Radu
  Florian}.} \bibinfo{year}{2016}\natexlab{}.
\newblock \showarticletitle{Toward Mention Detection Robustness with Recurrent
  Neural Networks.}
\newblock \bibinfo{journal}{\emph{CoRR}}  \bibinfo{volume}{abs/1602.07749}
  (\bibinfo{year}{2016}).
\newblock
\urldef\tempurl%
\url{http://dblp.uni-trier.de/db/journals/corr/corr1602.html#NguyenSDF16}
\showURL{%
\tempurl}


\bibitem[\protect\citeauthoryear{OECD}{OECD}{1996}]%
        {knowledgeoecd}
\bibfield{author}{\bibinfo{person}{OECD}.} \bibinfo{year}{1996}\natexlab{}.
\newblock \bibinfo{booktitle}{\emph{THE KNOWLEDGE-BASED ECONOMY. The
  Organisation for Economic Co-operation and Development}}.
\newblock
\urldef\tempurl%
\url{https://www.oecd.org/officialdocuments/publicdisplaydocumentpdf/?cote=OCDE/GD%2896%29102&docLanguage=En}
\showURL{%
\tempurl}


\bibitem[\protect\citeauthoryear{Oñoro-Rubio, Niepert, García-Durán,
  González-Sánchez, and López-Sastre}{Oñoro-Rubio et~al\mbox{.}}{2018}]%
        {onoro-rubio2018answering}
\bibfield{author}{\bibinfo{person}{Daniel Oñoro-Rubio},
  \bibinfo{person}{Mathias Niepert}, \bibinfo{person}{Alberto García-Durán},
  \bibinfo{person}{Roberto González-Sánchez}, {and}
  \bibinfo{person}{Roberto~J. López-Sastre}.} \bibinfo{year}{2018}\natexlab{}.
\newblock \showarticletitle{Answering Visual-Relational Queries in
  Web-Extracted Knowledge Graphs}. In \bibinfo{booktitle}{\emph{AKBC 2019}}.
\newblock


\bibitem[\protect\citeauthoryear{Pan, Dang, Yang, and Guo}{Pan
  et~al\mbox{.}}{2019}]%
        {chunguangel}
\bibfield{author}{\bibinfo{person}{Chunguang Pan}, \bibinfo{person}{Jingming
  Dang}, \bibinfo{person}{Zhi Yang}, {and} \bibinfo{person}{Xuyang Guo}.}
  \bibinfo{year}{2019}\natexlab{}.
\newblock \bibinfo{booktitle}{\emph{CCKS\&Baidu 2019 Chinese short-text entity
  linking (the first solution)}}.
\newblock
\urldef\tempurl%
\url{https://github.com/panchunguang/ccks_baidu_entity_link}
\showURL{%
\tempurl}


\bibitem[\protect\citeauthoryear{Qiu, Sun, Xu, Shao, Dai, and Huang}{Qiu
  et~al\mbox{.}}{2020}]%
        {DBLP:journals/corr/abs-2003-08271}
\bibfield{author}{\bibinfo{person}{Xipeng Qiu}, \bibinfo{person}{Tianxiang
  Sun}, \bibinfo{person}{Yige Xu}, \bibinfo{person}{Yunfan Shao},
  \bibinfo{person}{Ning Dai}, {and} \bibinfo{person}{Xuanjing Huang}.}
  \bibinfo{year}{2020}\natexlab{}.
\newblock \showarticletitle{Pre-trained Models for Natural Language Processing:
  {A} Survey}.
\newblock \bibinfo{journal}{\emph{arXiv preprint}}  \bibinfo{volume}{arXiv:
  2003.08271} (\bibinfo{year}{2020}).
\newblock


\bibitem[\protect\citeauthoryear{Ramesh, Dhariwal, Nichol, Chu, and
  Chen}{Ramesh et~al\mbox{.}}{2022}]%
        {DBLP:journals/corr/abs-2204-06125}
\bibfield{author}{\bibinfo{person}{Aditya Ramesh}, \bibinfo{person}{Prafulla
  Dhariwal}, \bibinfo{person}{Alex Nichol}, \bibinfo{person}{Casey Chu}, {and}
  \bibinfo{person}{Mark Chen}.} \bibinfo{year}{2022}\natexlab{}.
\newblock \showarticletitle{Hierarchical Text-Conditional Image Generation with
  {CLIP} Latents}.
\newblock \bibinfo{journal}{\emph{CoRR}}  \bibinfo{volume}{abs/2204.06125}
  (\bibinfo{year}{2022}).
\newblock


\bibitem[\protect\citeauthoryear{Schank and Abelson}{Schank and
  Abelson}{1975}]%
        {DBLP:conf/ijcai/SchenkA75}
\bibfield{author}{\bibinfo{person}{Roger~C. Schank} {and}
  \bibinfo{person}{Robert~P. Abelson}.} \bibinfo{year}{1975}\natexlab{}.
\newblock \showarticletitle{Scripts, Plans and Knowledge}. In
  \bibinfo{booktitle}{\emph{IJCAI}}. \bibinfo{pages}{151--157}.
\newblock
\urldef\tempurl%
\url{http://ijcai.org/Proceedings/75/Papers/021.pdf}
\showURL{%
\tempurl}


\bibitem[\protect\citeauthoryear{Sekine and Nobata}{Sekine and Nobata}{2004}]%
        {conf/lrec/SekineN04}
\bibfield{author}{\bibinfo{person}{Satoshi Sekine} {and}
  \bibinfo{person}{Chikashi Nobata}.} \bibinfo{year}{2004}\natexlab{}.
\newblock \showarticletitle{Definition, Dictionaries and Tagger for Extended
  Named Entity Hierarchy.}. In \bibinfo{booktitle}{\emph{LREC}}.
  \bibinfo{publisher}{European Language Resources Association}.
\newblock
\urldef\tempurl%
\url{http://dblp.uni-trier.de/db/conf/lrec/lrec2004.html#SekineN04}
\showURL{%
\tempurl}


\bibitem[\protect\citeauthoryear{Shuo, Jun, Jianzhuang, Qi, and Meng}{Shuo
  et~al\mbox{.}}{2020}]%
        {wang2020large}
\bibfield{author}{\bibinfo{person}{Wang Shuo}, \bibinfo{person}{Yue Jun},
  \bibinfo{person}{Liu Jianzhuang}, \bibinfo{person}{Tian Qi}, {and}
  \bibinfo{person}{Wang Meng}.} \bibinfo{year}{2020}\natexlab{}.
\newblock \showarticletitle{Large-scale few-shot learning via multi-modal
  knowledge discovery}. In \bibinfo{booktitle}{\emph{European Conference on
  Computer Vision}}. \bibinfo{pages}{718--734}.
\newblock


\bibitem[\protect\citeauthoryear{Strubell, Verga, Belanger, and
  McCallum}{Strubell et~al\mbox{.}}{2017}]%
        {conf/emnlp/StrubellVBM17}
\bibfield{author}{\bibinfo{person}{Emma Strubell}, \bibinfo{person}{Patrick
  Verga}, \bibinfo{person}{David Belanger}, {and} \bibinfo{person}{Andrew
  McCallum}.} \bibinfo{year}{2017}\natexlab{}.
\newblock \showarticletitle{Fast and Accurate Entity Recognition with Iterated
  Dilated Convolutions.}. In \bibinfo{booktitle}{\emph{EMNLP}},
  \bibfield{editor}{\bibinfo{person}{Martha Palmer}, \bibinfo{person}{Rebecca
  Hwa}, {and} \bibinfo{person}{Sebastian Riedel}} (Eds.).
  \bibinfo{publisher}{Association for Computational Linguistics},
  \bibinfo{pages}{2670--2680}.
\newblock
\showISBNx{978-1-945626-83-8}
\urldef\tempurl%
\url{http://dblp.uni-trier.de/db/conf/emnlp/emnlp2017.html#StrubellVBM17}
\showURL{%
\tempurl}


\bibitem[\protect\citeauthoryear{Suchanek, Kasneci, and Weikum}{Suchanek
  et~al\mbox{.}}{2007}]%
        {suchanek2007semantic}
\bibfield{author}{\bibinfo{person}{Fabian~M. Suchanek},
  \bibinfo{person}{Gjergji Kasneci}, {and} \bibinfo{person}{Gerhard Weikum}.}
  \bibinfo{year}{2007}\natexlab{}.
\newblock \showarticletitle{Yago: A Core of Semantic Knowledge}. In
  \bibinfo{booktitle}{\emph{Proceedings of the 16th International Conference on
  World Wide Web}} (Banff, Alberta, Canada) \emph{(\bibinfo{series}{WWW '07})}.
  \bibinfo{publisher}{ACM}, \bibinfo{address}{New York, NY, USA},
  \bibinfo{pages}{697--706}.
\newblock
\showISBNx{978-1-59593-654-7}
\urldef\tempurl%
\url{https://doi.org/10.1145/1242572.1242667}
\showDOI{\tempurl}


\bibitem[\protect\citeauthoryear{Sun, Cao, Zhao, Wan, Zhou, Zhang, Wang, and
  Zheng}{Sun et~al\mbox{.}}{2020}]%
        {sun2020multi}
\bibfield{author}{\bibinfo{person}{Rui Sun}, \bibinfo{person}{Xuezhi Cao},
  \bibinfo{person}{Yan Zhao}, \bibinfo{person}{Junchen Wan},
  \bibinfo{person}{Kun Zhou}, \bibinfo{person}{Fuzheng Zhang},
  \bibinfo{person}{Zhongyuan Wang}, {and} \bibinfo{person}{Kai Zheng}.}
  \bibinfo{year}{2020}\natexlab{}.
\newblock \showarticletitle{Multi-modal Knowledge Graphs for Recommender
  Systems}. In \bibinfo{booktitle}{\emph{Proceedings of the 29th ACM
  International Conference on Information and Knowledge Management}}. ACM,
  \bibinfo{pages}{1405--1414}.
\newblock


\bibitem[\protect\citeauthoryear{Sun, Zheng, Hao, and Qiu}{Sun
  et~al\mbox{.}}{2022}]%
        {sun-etal-2022-nsp}
\bibfield{author}{\bibinfo{person}{Yi Sun}, \bibinfo{person}{Yu Zheng},
  \bibinfo{person}{Chao Hao}, {and} \bibinfo{person}{Hangping Qiu}.}
  \bibinfo{year}{2022}\natexlab{}.
\newblock \showarticletitle{{NSP}-{BERT}: A Prompt-based Few-Shot Learner
  through an Original Pre-training Task {---}{---} Next Sentence Prediction}.
  In \bibinfo{booktitle}{\emph{Proceedings of the 29th International Conference
  on Computational Linguistics}}. \bibinfo{publisher}{International Committee
  on Computational Linguistics}, \bibinfo{address}{Gyeongju, Republic of
  Korea}, \bibinfo{pages}{3233--3250}.
\newblock
\urldef\tempurl%
\url{https://aclanthology.org/2022.coling-1.286}
\showURL{%
\tempurl}


\bibitem[\protect\citeauthoryear{Vrande{\v{c}}i{\'c} and
  Kr{\"o}tzsch}{Vrande{\v{c}}i{\'c} and Kr{\"o}tzsch}{2014}]%
        {VrandecicKroetzsch14cacm}
\bibfield{author}{\bibinfo{person}{Denny Vrande{\v{c}}i{\'c}} {and}
  \bibinfo{person}{Markus Kr{\"o}tzsch}.} \bibinfo{year}{2014}\natexlab{}.
\newblock \showarticletitle{Wikidata: A Free Collaborative Knowledgebase}.
\newblock \bibinfo{journal}{\emph{Commun. ACM}} \bibinfo{volume}{57},
  \bibinfo{number}{10} (\bibinfo{date}{Sep} \bibinfo{year}{2014}),
  \bibinfo{pages}{78--85}.
\newblock
\showISSN{0001-0782}
\urldef\tempurl%
\url{https://doi.org/10.1145/2629489}
\showDOI{\tempurl}


\bibitem[\protect\citeauthoryear{Wang, Qi, Wang, and Zheng}{Wang
  et~al\mbox{.}}{2019}]%
        {DBLP:conf/jist/WangQWZ19}
\bibfield{author}{\bibinfo{person}{Meng Wang}, \bibinfo{person}{Guilin Qi},
  \bibinfo{person}{Haofen Wang}, {and} \bibinfo{person}{Qiushuo Zheng}.}
  \bibinfo{year}{2019}\natexlab{}.
\newblock \showarticletitle{Richpedia: {A} Comprehensive Multi-modal Knowledge
  Graph}. In \bibinfo{booktitle}{\emph{Semantic Technology - 9th Joint
  International Conference, {JIST} 2019, Hangzhou, China, November 25-27, 2019,
  Proceedings}} \emph{(\bibinfo{series}{Lecture Notes in Computer Science},
  Vol.~\bibinfo{volume}{12032})}, \bibfield{editor}{\bibinfo{person}{Xin Wang},
  \bibinfo{person}{Francesca~Alessandra Lisi}, \bibinfo{person}{Guohui Xiao},
  {and} \bibinfo{person}{Elena Botoeva}} (Eds.). \bibinfo{publisher}{Springer},
  \bibinfo{pages}{130--145}.
\newblock


\bibitem[\protect\citeauthoryear{Wang, Wang, Qi, and Zheng}{Wang
  et~al\mbox{.}}{2020}]%
        {DBLP:journals/bdr/WangWQZ20}
\bibfield{author}{\bibinfo{person}{Meng Wang}, \bibinfo{person}{Haofen Wang},
  \bibinfo{person}{Guilin Qi}, {and} \bibinfo{person}{Qiushuo Zheng}.}
  \bibinfo{year}{2020}\natexlab{}.
\newblock \showarticletitle{Richpedia: {A} Large-Scale, Comprehensive
  Multi-Modal Knowledge Graph}.
\newblock \bibinfo{journal}{\emph{Big Data Res.}}  \bibinfo{volume}{22}
  (\bibinfo{year}{2020}), \bibinfo{pages}{100159}.
\newblock
\urldef\tempurl%
\url{https://doi.org/10.1016/j.bdr.2020.100159}
\showDOI{\tempurl}


\bibitem[\protect\citeauthoryear{Wu, Jiang, Lei, and Xu}{Wu
  et~al\mbox{.}}{2015}]%
        {conf/medinfo/WuJLX15}
\bibfield{author}{\bibinfo{person}{Yonghui Wu}, \bibinfo{person}{Min Jiang},
  \bibinfo{person}{Jianbo Lei}, {and} \bibinfo{person}{Hua Xu}.}
  \bibinfo{year}{2015}\natexlab{}.
\newblock \showarticletitle{Named Entity Recognition in Chinese Clinical Text
  Using Deep Neural Network.}. In \bibinfo{booktitle}{\emph{MedInfo}}
  \emph{(\bibinfo{series}{Studies in Health Technology and Informatics},
  Vol.~\bibinfo{volume}{216})}, \bibfield{editor}{\bibinfo{person}{Indra~Neil
  Sarkar}, \bibinfo{person}{Andrew Georgiou}, {and}
  \bibinfo{person}{Paulo~Mazzoncini de~Azevedo~Marques}} (Eds.).
  \bibinfo{publisher}{IOS Press}, \bibinfo{pages}{624--628}.
\newblock
\showISBNx{978-1-61499-564-7}
\urldef\tempurl%
\url{http://dblp.uni-trier.de/db/conf/medinfo/medinfo2015.html#WuJLX15}
\showURL{%
\tempurl}


\bibitem[\protect\citeauthoryear{Xinlei, Shrivastava, and Gupta}{Xinlei
  et~al\mbox{.}}{2013}]%
        {chen2013neil}
\bibfield{author}{\bibinfo{person}{Chen Xinlei}, \bibinfo{person}{Abhinav
  Shrivastava}, {and} \bibinfo{person}{Abhinav Gupta}.}
  \bibinfo{year}{2013}\natexlab{}.
\newblock \showarticletitle{Neil: Extracting visual knowledge from web data}.
  In \bibinfo{booktitle}{\emph{Proceedings of the IEEE International Conference
  on Computer Vision}}. \bibinfo{pages}{1409--1416}.
\newblock


\bibitem[\protect\citeauthoryear{Xu, Xu, Liang, Xie, Liang, Cui, and Xiao}{Xu
  et~al\mbox{.}}{2017}]%
        {DBLP:conf/ieaaie/XuXLXLCX17}
\bibfield{author}{\bibinfo{person}{Bo Xu}, \bibinfo{person}{Yong Xu},
  \bibinfo{person}{Jiaqing Liang}, \bibinfo{person}{Chenhao Xie},
  \bibinfo{person}{Bin Liang}, \bibinfo{person}{Wanyun Cui}, {and}
  \bibinfo{person}{Yanghua Xiao}.} \bibinfo{year}{2017}\natexlab{}.
\newblock \showarticletitle{CN-DBpedia: {A} Never-Ending Chinese Knowledge
  Extraction System}. In \bibinfo{booktitle}{\emph{Advances in Artificial
  Intelligence: From Theory to Practice - 30th International Conference on
  Industrial Engineering and Other Applications of Applied Intelligent Systems,
  {IEA/AIE} 2017, Arras, France, June 27-30, 2017, Proceedings, Part {II}}}
  \emph{(\bibinfo{series}{Lecture Notes in Computer Science},
  Vol.~\bibinfo{volume}{10351})}, \bibfield{editor}{\bibinfo{person}{Salem
  Benferhat}, \bibinfo{person}{Karim Tabia}, {and} \bibinfo{person}{Moonis
  Ali}} (Eds.). \bibinfo{publisher}{Springer}, \bibinfo{pages}{428--438}.
\newblock
\urldef\tempurl%
\url{https://doi.org/10.1007/978-3-319-60045-1\_44}
\showDOI{\tempurl}


\bibitem[\protect\citeauthoryear{Yamada, Washio, Shindo, and Matsumoto}{Yamada
  et~al\mbox{.}}{2020}]%
        {Yamada2020}
\bibfield{author}{\bibinfo{person}{Ikuya Yamada}, \bibinfo{person}{Koki
  Washio}, \bibinfo{person}{Hiroyuki Shindo}, {and} \bibinfo{person}{Yuji
  Matsumoto}.} \bibinfo{year}{2020}\natexlab{}.
\newblock \showarticletitle{Global Entity Disambiguation with Pretrained
  Contextualized Embeddings of Words and Entities}.
\newblock \bibinfo{journal}{\emph{arXiv preprint arXiv:1909.00426}}
  (\bibinfo{year}{2020}).
\newblock


\bibitem[\protect\citeauthoryear{Yang, Gu, Lin, Tang, Zhuang, Wu, Chen, Hu, and
  Ren}{Yang et~al\mbox{.}}{2019}]%
        {Yang2019}
\bibfield{author}{\bibinfo{person}{Xiyuan Yang}, \bibinfo{person}{Xiaotao Gu},
  \bibinfo{person}{Sheng Lin}, \bibinfo{person}{Siliang Tang},
  \bibinfo{person}{Yueting Zhuang}, \bibinfo{person}{Fei Wu},
  \bibinfo{person}{Zhigang Chen}, \bibinfo{person}{Guoping Hu}, {and}
  \bibinfo{person}{Xiang Ren}.} \bibinfo{year}{2019}\natexlab{}.
\newblock \showarticletitle{Learning dynamic context augmentation for global
  entity linking}. In \bibinfo{booktitle}{\emph{Proceedings of the 2019
  Conference on Empirical Methods in Natural Language Processing and the
  International Joint Conference on Natural Language Processing (EMNLP-IJCNLP
  2019)}}. \bibinfo{pages}{271--281}.
\newblock


\bibitem[\protect\citeauthoryear{Yu, Jiang, Yang, and Xia}{Yu
  et~al\mbox{.}}{2020}]%
        {yu-etal-2020-improving-multimodal}
\bibfield{author}{\bibinfo{person}{Jianfei Yu}, \bibinfo{person}{Jing Jiang},
  \bibinfo{person}{Li Yang}, {and} \bibinfo{person}{Rui Xia}.}
  \bibinfo{year}{2020}\natexlab{}.
\newblock \showarticletitle{Improving Multimodal Named Entity Recognition via
  Entity Span Detection with Unified Multimodal Transformer}. In
  \bibinfo{booktitle}{\emph{Proceedings of the 58th Annual Meeting of the
  Association for Computational Linguistics}}. \bibinfo{publisher}{Association
  for Computational Linguistics}, \bibinfo{address}{Online},
  \bibinfo{pages}{3342--3352}.
\newblock
\urldef\tempurl%
\url{https://doi.org/10.18653/v1/2020.acl-main.306}
\showDOI{\tempurl}


\bibitem[\protect\citeauthoryear{Zellers, Lu, Hessel, Yu, Park, Cao, Farhadi,
  and Choi}{Zellers et~al\mbox{.}}{2021}]%
        {DBLP:conf/nips/ZellersLHYPCFC21}
\bibfield{author}{\bibinfo{person}{Rowan Zellers}, \bibinfo{person}{Ximing Lu},
  \bibinfo{person}{Jack Hessel}, \bibinfo{person}{Youngjae Yu},
  \bibinfo{person}{Jae~Sung Park}, \bibinfo{person}{Jize Cao},
  \bibinfo{person}{Ali Farhadi}, {and} \bibinfo{person}{Yejin Choi}.}
  \bibinfo{year}{2021}\natexlab{}.
\newblock \showarticletitle{{MERLOT:} Multimodal Neural Script Knowledge
  Models}. In \bibinfo{booktitle}{\emph{Advances in Neural Information
  Processing Systems 34: Annual Conference on Neural Information Processing
  Systems 2021, NeurIPS 2021, December 6-14, 2021, virtual}},
  \bibfield{editor}{\bibinfo{person}{Marc'Aurelio Ranzato},
  \bibinfo{person}{Alina Beygelzimer}, \bibinfo{person}{Yann~N. Dauphin},
  \bibinfo{person}{Percy Liang}, {and} \bibinfo{person}{Jennifer~Wortman
  Vaughan}} (Eds.). \bibinfo{pages}{23634--23651}.
\newblock
\urldef\tempurl%
\url{https://proceedings.neurips.cc/paper/2021/hash/c6d4eb15f1e84a36eff58eca3627c82e-Abstract.html}
\showURL{%
\tempurl}


\bibitem[\protect\citeauthoryear{Zhang and Elhadad}{Zhang and Elhadad}{2013}]%
        {journals/jbi/ZhangE13}
\bibfield{author}{\bibinfo{person}{Shaodian Zhang} {and}
  \bibinfo{person}{Noemie Elhadad}.} \bibinfo{year}{2013}\natexlab{}.
\newblock \showarticletitle{Unsupervised biomedical named entity recognition:
  Experiments with clinical and biological texts.}
\newblock \bibinfo{journal}{\emph{J. Biomed. Informatics}}
  \bibinfo{volume}{46}, \bibinfo{number}{6} (\bibinfo{year}{2013}),
  \bibinfo{pages}{1088--1098}.
\newblock
\urldef\tempurl%
\url{http://dblp.uni-trier.de/db/journals/jbi/jbi46.html#ZhangE13}
\showURL{%
\tempurl}


\bibitem[\protect\citeauthoryear{Zhang}{Zhang}{2020}]%
        {Zhang2020}
\bibfield{author}{\bibinfo{person}{Tongxi Zhang}.}
  \bibinfo{year}{2020}\natexlab{}.
\newblock \showarticletitle{A Brief Study on Short Video Platform and
  Education}. In \bibinfo{booktitle}{\emph{Proceedings of the 2nd International
  Conference on Literature, Art and Human Development (ICLAHD 2020)}}.
  \bibinfo{publisher}{Atlantis Press}, \bibinfo{pages}{543--547}.
\newblock
\showISBNx{978-94-6239-304-2}
\showISSN{2352-5398}
\urldef\tempurl%
\url{https://doi.org/10.2991/assehr.k.201215.494}
\showDOI{\tempurl}


\bibitem[\protect\citeauthoryear{Zheng, Wang, Bao, Hao, Zhou, and Xu}{Zheng
  et~al\mbox{.}}{2017}]%
        {conf/acl/ZhengWBHZX17}
\bibfield{author}{\bibinfo{person}{Suncong Zheng}, \bibinfo{person}{Feng Wang},
  \bibinfo{person}{Hongyun Bao}, \bibinfo{person}{Yuexing Hao},
  \bibinfo{person}{Peng Zhou}, {and} \bibinfo{person}{Bo Xu}.}
  \bibinfo{year}{2017}\natexlab{}.
\newblock \showarticletitle{Joint Extraction of Entities and Relations Based on
  a Novel Tagging Scheme.}. In \bibinfo{booktitle}{\emph{ACL (1)}},
  \bibfield{editor}{\bibinfo{person}{Regina Barzilay} {and}
  \bibinfo{person}{Min-Yen Kan}} (Eds.). \bibinfo{publisher}{Association for
  Computational Linguistics}, \bibinfo{pages}{1227--1236}.
\newblock
\showISBNx{978-1-945626-75-3}
\urldef\tempurl%
\url{http://dblp.uni-trier.de/db/conf/acl/acl2017-1.html#ZhengWBHZX17}
\showURL{%
\tempurl}


\bibitem[\protect\citeauthoryear{Zhou, Zheng, Xu, Qi, Bao, and Xu}{Zhou
  et~al\mbox{.}}{2017}]%
        {conf/cncl/ZhouZXQBX17}
\bibfield{author}{\bibinfo{person}{Peng Zhou}, \bibinfo{person}{Suncong Zheng},
  \bibinfo{person}{Jiaming Xu}, \bibinfo{person}{Zhenyu Qi},
  \bibinfo{person}{Hongyun Bao}, {and} \bibinfo{person}{Bo Xu}.}
  \bibinfo{year}{2017}\natexlab{}.
\newblock \showarticletitle{Joint Extraction of Multiple Relations and Entities
  by Using a Hybrid Neural Network.}. In \bibinfo{booktitle}{\emph{CCL}}
  \emph{(\bibinfo{series}{Lecture Notes in Computer Science},
  Vol.~\bibinfo{volume}{10565})}, \bibfield{editor}{\bibinfo{person}{Maosong
  Sun}, \bibinfo{person}{Xiaojie Wang}, \bibinfo{person}{Baobao Chang}, {and}
  \bibinfo{person}{Deyi Xiong}} (Eds.). \bibinfo{publisher}{Springer},
  \bibinfo{pages}{135--146}.
\newblock
\showISBNx{978-3-319-69005-6}
\urldef\tempurl%
\url{http://dblp.uni-trier.de/db/conf/cncl/ccl2017.html#ZhouZXQBX17}
\showURL{%
\tempurl}


\bibitem[\protect\citeauthoryear{Zhu, Li, Wang, Jiang, Sun, Wang, Xiao, and
  Yuan}{Zhu et~al\mbox{.}}{2022}]%
        {DBLP:journals/corr/abs-2202-05786}
\bibfield{author}{\bibinfo{person}{Xiangru Zhu}, \bibinfo{person}{Zhixu Li},
  \bibinfo{person}{Xiaodan Wang}, \bibinfo{person}{Xueyao Jiang},
  \bibinfo{person}{Penglei Sun}, \bibinfo{person}{Xuwu Wang},
  \bibinfo{person}{Yanghua Xiao}, {and} \bibinfo{person}{Nicholas~Jing Yuan}.}
  \bibinfo{year}{2022}\natexlab{}.
\newblock \showarticletitle{Multi-Modal Knowledge Graph Construction and
  Application: {A} Survey}.
\newblock \bibinfo{journal}{\emph{CoRR}}  \bibinfo{volume}{abs/2202.05786}
  (\bibinfo{year}{2022}).
\newblock
\showeprint[arXiv]{2202.05786}
\urldef\tempurl%
\url{https://arxiv.org/abs/2202.05786}
\showURL{%
\tempurl}


\end{thebibliography}

\end{document}